\documentclass[a4paper,11pt]{article}
\pdfoutput=1 

\usepackage{jheppub} 

\usepackage[T1]{fontenc} 

\providecommand{\abs}[1]{\lvert#1\rvert}

\title{\boldmath U-duality in quantum M2-branes and gauged supergravities}

 \author[a]{M.P. García del Moral,}
 \author[b]{C. las Heras,}
 \author[c]{A. Restuccia,}
\affiliation[a]{Universidad de la Rioja,\\Área de Física, Departamento de Química, Centro Científico Tecnológico Universidad de la Rioja, La Rioja, 26006, Spain}
 \affiliation[b]{Instituto de Física Teórica UAM/CSIC,\\ C/ Nicolás Cabrera 13-15 \\ Universidad Autónoma de Madrid \\ Cantoblanco, Madrid 28040, Spain}
 
 \affiliation[c]{Universidad de Antofagasta,\\Departamento de Física, Universidad de Antofagasta, Aptdo 02800, Chile}
\emailAdd{m-pilar.garciam@unirioja.es}
\emailAdd{camilo.lasheras@ift.csic.es}
\emailAdd{alvaro.restuccia@uantof.cl}

\abstract{In this paper, we study the relation of the M2-brane with fluxes and monodromy in $SL(2,\mathbb{Z})$, which has a quantum discrete supersymmetric spectrum with finite multiplicity, and type IIB gauged supergravities in nine dimensions. $SL(2,\mathbb{Z})$ is the group of isotopy classes of the area preserving diffeomorphisms. The global description of these M2-branes we are considering is formulated on twisted torus bundles, and they are classified in terms of $H^2(\Sigma,\mathcal{Z}_{\mathcal{\rho}})$, or equivalently, by their coinvariants for a given monodromy subgroup. We find the 'gauge' symmetries between equivalent M2-branes, on torus bundles with monodromy, that leads to $\mathbb{R}$, $SO(2)$ or $SO(1,1)$, the symmetry groups of type IIB gauged supergravities in 9d. We obtain an explicit relation between the equivalent classes of M2-brane bundles and the mass parameters that classify the gaugings of type IIB supergravities in 9d. We also find that the symmetries, between inequivalent M2-branes on twisted torus bundles for a given monodromy, are related with $\mathbb{Z}$, $\mathcal{Z}_3$, $\mathcal{Z}_5$, $\mathcal{Z}_9$ or $\mathcal{Z}_{2n-7}$ for $n\geq 5$, the U-duality symmetry group, a subgroup of $SL(2,\mathbb{Z})$. In distinction, in the case without monodromy, related to type II maximal supergravity at low energies, its U-duality group corresponds to the full $SL(2,\mathbb{Z})$. 
}

\begin{document}
\emergencystretch 3em
\hypersetup{pageanchor=false}
\makeatletter
\let\old@fpheader\@fpheader
\preprint{IFT-UAM/CSIC-24-68}

\makeatother
\maketitle
\flushbottom

\section{Introduction}
\label{sec:intro}
In this paper we study the relationship between the M-theory description, type II gauged supergravities, and U-dualities. To this end, we study the M2-brane microscopic description compactified on a simple but still illustrative target space $M_9\times T^2$ with $C_{\pm}$ fluxes and we compare it with the maximal and gauged type IIB supergravities in nine dimensions.

Type II supergravities in nine dimensions are well known and characterized (See for example \cite{Bergshoeff5,Bergshoeff9,Ortin,Meessen,Melgarejo,Hull4,Hull8}). There is a unique maximal supergravity, that can be obtained by KK reduction from type IIA or type IIB supergravity in $D=10$ and eight massive deformations. Four are in the type IIA sector and the other four on the type IIB. Type II gauged supergravities in nine dimensions can be obtained by dimensional reduction (Scherk Schwarz of Kaluza Klein) from higher dimensional theories or by deforming the massless theory on nine dimensions (via the embedding tensor formalism) \cite{Samtleben,Samtleben2,Samtleben4,Samtleben6}. 

On massive type II supergravities in nine dimensions, it occurs that different subgroups of the global symmetry group are promoted to local symmetries. The degrees of freedom are the same in all cases, but their corresponding lagrangians acquire new terms related to the coupling with the gauge vectors (covariant derivates), to the scalar potential and imply a modification of some of the field strengths. Besides the eight massive deformations, where the mass parameter is unique in each case, other supergravities may arise as combinations of the previous ones \cite{Bergshoeff5,Bergshoeff9}). Domain wall solutions from all these type II supergravities have been explored in several papers \cite{Bergshoeff5,Bergshoeff9,Cowdall}.

Regarding the quantum description of type II supergravities in nine dimensions, the most well-known case corresponds to the maximal supergravity, which arises as the low energy limit of type II string theories on a circle. On the other hand, the UV description of type II gauge counterparts was given in \cite{Hull4,Hull8} (for the type IIB side) in terms of Scherk Schwarz reduction of type IIB string theory in ten dimensions. 

From the M-theory perspective, we know that the continuous spectra of the M2-branes \cite{deWit6} imply that the theory must be understood as a second quantization theory. The main reason was the existence of string configurations with vanishing costs in energy that are responsible for generating instabilities in the theory. Nevertheless, there are sectors of M2-branes where there do not exist those types of string-like configurations that are stable at supersymmetric quantum level. The first one of these sectors was found in \cite{Restuccia} on a toroidal background with a topological restriction imposed on the embedding maps of the theory, it implies the existence of a nontrivial flux condition on the worldvolume of the M2-brane. It was named central charge condition because it appears as a central charge term on the superalgebra of the theory. It was rigorously proven that its $SU(N)$ regularized model has a discrete supersymmetric spectrum \cite{Boulton}. Consequently, it may describe microscopical degrees of freedom of, at least, a sector of M-theory. Other sectors of M2-branes with a good quantum description were found in \cite{mpgm15,mpgm11,mpgm6,mpgm24}. In \cite{mpgm2,mpgm17,mpgm23} it was claimed that all type II gauged supergravities in nine dimensions are related to particular sectors of the M2-brane on a torus with a discrete supersymmetric spectrum.

This work is organized as follows: In section 2, we will review type IIB supergravity in ten and nine dimensions, emphasizing the case where the mass parameters transform as a triplet of $SL(2,\mathbb{R})$. In section 3, the worldvolume description of the M2-branes with nonvanishing winding on a torus is analyzed, together with it global description in terms of twisted torus bundles with monodromy in $SL(2,\mathbb{Z})$. In section 4, we present our main results, we show that the group transformation between equivalent twisted torus bundles for a given flux and monodromy, correspond with the restricted gauge symmetries at low energies. This allows to give a $D=11$ interpretation to the discrete values the mass parameters take in the quantum description. On the other hand, the group transformation between inequivalent twisted torus bundles corresponds to the U-duality group, corresponding to subgroups of $SL(2,\mathbb{Z})$ for the gauged/massive cases.

\section{Type IIB supergravity}

In this section, we will briefly review the description of type IIB supergravities in nine and ten dimensions. We will emphasize the type IIB gauged supergravities in $D=9$ \cite{Bergshoeff5,Hull4,Hull8,Melgarejo}, following from a Scherk Schwarz reduction of higher-dimensional theories or by deformations of maximal supergravity on the same dimension via embedding tensor formalism. We will restrict ourselves to the case where the mass parameters transform as a triplet, which corresponds to the gauging of the $\mathbb{R}$, $SO(2)$, or $SO(1,1)$ subgroups of $SL(2,\mathbb{R})$. Such reduction was considered in \cite{Bergshoeff8,Pope,Kaloper2} for particular elements of $SL(2,\mathbb{R})$ and in \cite{Ortin, Meessen,Cowdall} for general elements. We will also mention their quantum descriptions in terms of type IIB string theory and F-theory.
\subsection{Type IIB supergravity in $D=10$}
It is known that the action of $N=2$ type IIB supergravity in ten dimensions can be written in a $SL(2,\mathbb{R})$ covariant form \cite{Schwarz6}. The field content is given by 
\begin{eqnarray}
    \left\lbrace \hat{e}_{\hat{\mu}}^{\hat{a}}, \hat{\phi}, \hat{\chi}, \hat{B}_{\hat{\mu}\hat{\nu}}^{(1)},\hat{B}_{\hat{\mu}\hat{\nu}}^{(2)},\hat{D}_{\hat{\mu}\hat{\nu}\hat{\rho}\hat{\sigma}}, \hat{\psi}_{\hat{\mu}}, \hat{\lambda} \right\rbrace
\end{eqnarray}
and they transform under the global $SL(2,\mathbb{R})$ symmetry as \cite{Hull4}
\begin{eqnarray}
     \hat{\tau} &\rightarrow& \frac{a\hat{\tau} + b}{c\hat{\tau}+d}, \, \hspace{2.5cm} \, \Vec{\hat{B}}\rightarrow \mathcal{M}\Vec{\hat{B}}, \, \hspace{2.5cm} \, \displaystyle \hat{D}\rightarrow \hat{D}, \nonumber \\
     \hat{\psi}_{\hat{\mu}} &\rightarrow& \left(\frac{c\hat{\tau*}+d}{c\hat{\tau}+d}\right)^{1/4}\hat{\psi}_{\hat{\mu}}, \, \hspace{0.7cm} \, \hat{\lambda}\rightarrow \left(\frac{c\hat{\tau*}+d}{c\hat{\tau}+d}\right)^{3/4} \hat{\lambda}, \, \hspace{0.7cm} \, \hat{\epsilon} \rightarrow  \left(\frac{c\hat{\tau*}+d}{c\hat{\tau}+d}\right)^{1/4}\hat{\epsilon} 
\end{eqnarray}
with $\hat{\tau}=\hat{\chi} + e^{-\hat{\phi}}$ the axio-dilaton, $\Vec{\hat{B}}= \left(\begin{array}{c}
     \hat{B}^{(1)} \\
     \hat{B}^{(2)}
\end{array}\right)$ and $\mathcal{M} = \left(\begin{array}{cc}
    a & b \\
     c & d
\end{array}\right)\in SL(2,\mathbb{R})$.

The elements of $SL(2,\mathbb{R})$ can be classified according to its trace in parabolic, elliptic, or hyperbolic elements for $\abs{Tr(\mathcal{M})}=2$, $\abs{Tr(\mathcal{M})}<2$ or $\abs{Tr(\mathcal{M})}>2$, respectively. This classification is based on subsets and does not generally correspond to subgroups. Nevertheless, as it is known (see appendix \ref{ApeA}) each matrix of any of these subsets is conjugate to a member of one of the following three one-parameter subgroups \cite{Hull4}:
\begin{enumerate}
    \item Non compact parabolic subgroup  $\mathbb{R}$ generated by
    \begin{eqnarray}
        \mathcal{M}_p= \left( \begin{array}{cc}
           1  & k \\
            0 & 1
        \end{array} \right) \label{parabolicmonodromy}
    \end{eqnarray}
with $k\in\mathbb{R}$. Each element defines a parabolic conjugacy class with $Tr(\mathcal{M})=2$.
\item A compact elliptic subgroup $SO(2)$ generated by
\begin{eqnarray}
    \mathcal{M}_e = \left( \begin{array}{cc}
      \cos(\theta)   &  \mp\sin(\theta) \\
        \pm\sin(\theta) &  \cos(\theta)
    \end{array}\right) \label{ellipticmonodromy}
\end{eqnarray}
with $\theta \in \left[0,2\pi \right]$. Each element defines a conjugacy class with $\abs{Tr(\mathcal{M})}<2$
\item A non compact hyperbolic $SO(1,1)^+$ subgroup generated by
\begin{eqnarray}
    \mathcal{M}_h = \left( \begin{array}{cc}
       e^a  & 0 \\
        0 & e^{-a}
    \end{array}\right) \label{hiperbolicmonodromy}
\end{eqnarray}
with $a\in \mathbb{R}$. Each element defines a conjugacy class with $Tr(\mathcal{M})>2$.     
\end{enumerate}

All these generators can be written via the exponential map as $\mathcal{M}=e^M$ where $M$ an element of the algebra, denoted as mass matrix
\begin{eqnarray}
    M= \left( \begin{array}{cc}
      m_1   & m_2+m_3 \\
        m_2 - m_3 &  -m_1
    \end{array} \right), \label{massmatrix}
\end{eqnarray}
Therefore, the parabolic case will correspond to $m_1=0$ and $m_2=m_3=\frac{a}{2}$, the elliptic case to $m_1=m_2=0$ and $m_3=\theta$ and the hyperbolic to $m_1=a$, $m_2=m_3=0$.
\subsection{Type IIB gauged supergravities in $D=9$}

\begin{figure}
    \centering
    \includegraphics[scale=0.9]{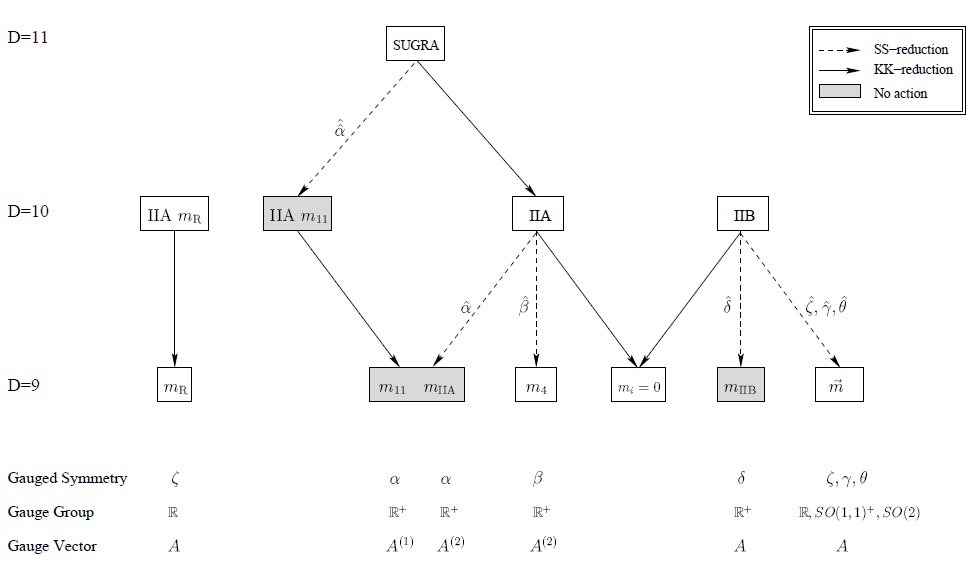}
    \caption{Type II supergravities in $D=9$ from \cite{Bergshoeff5}}
    \label{fig1}
\end{figure}

Type II supergravities in $D=9$ are well-known \cite{Bergshoeff5,Hull4,Hull8,Melgarejo}. There is a unique maximal supergravity (as it is shown in figure \ref{fig1})\footnote{See \cite{AbouZeid} to take into account subtleties in the massive sector to this statement.}, which can be obtained from KK reduction of any of the type II (IIA or IIB) supergravities in ten dimensions. Equivalently, it can be obtained by a toroidal compactification of supegravity in eleven dimensions. It field content is given by
\begin{eqnarray}
    \left\lbrace e_{\mu}^{a}, \phi,\varphi, \chi, A_\mu, A_\mu^{(1)}, A_\mu^{(2)},   B_{\mu\nu}^{(1)},B_{\mu\nu}^{(2)},C_{\mu\nu\rho}, \psi_{\mu}, \lambda, \tilde{\lambda} \right\rbrace
\end{eqnarray}
The massless theory in nine dimensions inherits global symmetries from the higher dimensional theory. These are: two $\mathbb{R}^+$ from type IIA supergravity and one $\mathbb{R}^+$ plus a full $SL(2,\mathbb{R})$ from type IIB supergravity.
The $SL(2,\mathbb{R})$ global symmetry in nine dimensions is given by
\begin{eqnarray}
     \tau&\rightarrow& \frac{a\tau + b}{c\tau+d}, \, \hspace{2.5cm} \, \Vec{B}\rightarrow \mathcal{M}\Vec{B}, \, \hspace{2.5cm} \, \Vec{A}\rightarrow \mathcal{M}\Vec{A}, \nonumber \\
    \psi_{\mu} &\rightarrow& \left(\frac{c\tau*+d}{c\tau+d}\right)^{1/4}\psi_{\mu}, \, \hspace{0.7cm} \, \lambda\rightarrow \left(\frac{c\tau*+d}{c\tau+d}\right)^{3/4} \lambda, \, \hspace{0.7cm} \, \hat{\epsilon} \rightarrow  \left(\frac{c\hat{\tau*}+d}{c\hat{\tau}+d}\right)^{1/4}\hat{\epsilon}, \nonumber \\
    \tilde{\lambda}   &\rightarrow& \left(\frac{c\tau*+d}{c\tau+d}\right)^{-1/4}\tilde{\lambda}, \, \hspace{1cm} \, \varphi \rightarrow \varphi , \, \hspace{3cm} \, C \rightarrow C
\end{eqnarray}
There are also local symmetries on the theory, i.e the transformations of the gauge vectors
\begin{eqnarray}
    A^{(1)} &\rightarrow& A^{(1)} - d\lambda^{(1)}, \hspace{0.5cm} A^{(2)} \rightarrow A^{(2)} - d\lambda^{(2)} \\
    A &\rightarrow& A - d\lambda, \hspace{1cm} \Vec{B} \rightarrow \Vec{B} - \Vec{A}d\lambda
\end{eqnarray}
Besides the maximal supergravity, there are also eight type II gauged supergravities in nine dimensions. Each one should correspond to a gauging of the global symmetry from the massless theory. These gauged (massive) supergravities were obtained by: performing a Scherk Schwarz (KK) reduction of a masless (massive) supergravity in ten dimension or by the deformation of the massless theory on the same dimension via the embedding tensor formalism.

Four out of eight type II gauged supergravities in nine dimensions are on the type IIA sector and the rest are on the type IIB associated with the gauging of the global symmetry $GL(2,\mathbb{R})=\mathbb{R}\times SL(2,\mathbb{R})$. The gauging of $\mathbb{R}$, corresponds to the gauging a trombone symmetry of the equations of motion that will not be considered any further in the present analysis. The other three of these type IIB reductions are related to the case where the mass parameters transform as a triplet of $SL(2,\mathbb{R})$. The gauging  corresponds to the gauging of the $\mathbb{R}$, $SO(2)$ or $SO(1,1)^+$ subgroups of $SL(2,\mathbb{R})$. Such reductions were considered in \cite{Bergshoeff8,Pope,Kaloper2} for particular elements and in \cite{Ortin, Meessen,Cowdall} for general elements of $SL(2,\mathbb{R})$.
The main idea of the Scherk Schwarz  mechanism \cite{Schwarz5} is to introduce mass parameters in toroidal compactifications by considering a twisted reduction of the higher dimensional fields such that they explicitly depend on the torus coordinate $y$ as 
\begin{eqnarray}
    \phi (x^\mu,y)=g_y(\phi^i(x^\mu)),
\end{eqnarray}
where $g_y=g(y)$ is a symmetry transformation contained in a subgroup of the global symmetry of the higher dimensional theory.

The map $g(y)$ is not periodic, but has a monodromy $\mathcal{M}(g)=\exp M$ contained in the symmetry group, where $M=g^{-1}\partial_y g$ is an element of the Lie algebra, independent of the torus coordinate and proportional to the mass matrix of the dimensionally reduced theory. A given monodromy may result in a different, but physically equivalent, mass matrices. Consequently, the reductions are classified by conjugacy classes of the monodromy $\mathcal{M}$. As a result, the global symmetry of the higher dimensional theory is broken to a subgroup in the reduced one.

For the type IIB supergravity, the global symmetry group is $SL(2,\mathbb{R})$ and there are three distinct classes of inequivalent theories corresponding to parabolic (\ref{parabolicmonodromy}), elliptic (\ref{ellipticmonodromy}) and hyperbolic (\ref{hiperbolicmonodromy}) conjugacy classes of monodromy. Each one is generated by mass matrices \cite{Hull4}
\begin{eqnarray}
    M_p = \left( \begin{array}{cc}
        0 & a \\
        0 & 0
    \end{array} \right)\, , \, M_e=\left( \begin{array}{cc}
        0 & \theta \\
        -\theta & 0
    \end{array} \right)\, , \, M_h=\left( \begin{array}{cc}
        a & 0 \\
        0 & -a
    \end{array} \right) \label{massmatrices}
\end{eqnarray}
as it can be seen from (\ref{massmatrix}). On (\ref{massmatrices}), both $a$'s are noncompact parameters while $\theta$ is the compact one.

The monodromy and the mass matrices will appear on the reduced theory modifying the Lagrangian, inducing the appearance of covariant derivatives and modifying the field strengths of the background fields.

On the other hand, the embedding tensor formalism is not a higher dimensional approach but a method to gauge the global symmetry group of the massless theory on the same dimension. Let us consider an algebra \cite{Samtleben2,Samtleben3,Samtleben4,Samtleben5,Samtleben6,Samtleben}
\begin{eqnarray}
    \left[t_i,t_j\right] = f_{ij}^k t_k
    \end{eqnarray}
where $t_i$ with $i=1,\dots,\mbox{dim}(\mathcal{G}_0)$ are the generators of the Lie algebra $\mathcal{G}_0$ of the global symmetry group $G_0$ and $f_{ij}^k$ the structure constants.

The embedding tensor formalism couples the gauge vector fields $A_\mu^M$ to the $G_0$ symmetry generators $t_\alpha$ in the form of a covariant derivative
\begin{eqnarray}
    D_\mu = \partial_\mu - g A_\mu^M\Theta_M^it_i
\end{eqnarray}
where $\mu$ is space-time index, $g$ is the gauge coupling constant, $M$ labelled the representation of the global symmetry group and $\Theta_M^i$ is the embedding tensor. The latter is a rectangular matrix that allows to couple the vector fields with the symmetry generators. The embedding tensor can be understood as a map from the representation of the global symmetry group to the gauge group of local symmetry. Indeed, $X_M=\Theta_M^it_i$ are interpreted as the generators of the gauge group that will appear on the gauge transformations of the fields. 

By considering a particular gauging, the embedding tensor breaks down the invariance under the global symmetry group to the local gauge group. In the particular case of type IIB supergravity, we have that the nonzero components of the embedding tensor are given by $\Theta_0^n=m_n$ with $n=1,2,3$ and $m_n$ the mass parameter of the matrix (\ref{massmatrix}).

Type IIB gauge supergravities in nine dimensions can be obtained by Scherk-Schwarz reduction from a higher dimensional theory, or by deforming the massless case via the embedding tensor formalism. Both methods leads to equivalent theories in each case.

\subsection{Scherk Schwarz reduction of type IIB superstring}

It is argued in \cite{Hull} that the symmetry group breaks to the discrete subgroup $SL(2,\mathbb{Z})$ in the quantum theory in order to guarantee the quantization of the charges. Therefore, the quantum consistent Scherk Schwarz reductions of type IIB superstring, are those for which the monodromy resides in the conjugacy classes of $SL(2,\mathbb{Z})$ \cite{Zwiebach,Zwiebach2}. It can be seen as a restriction on the choice of $M$ such that $\mathcal{M}$ is discrete. Consequently, the compact or non compact parameters of the subgroups of $SL(2,\mathbb{R})$ become restricted in the quantum theory.

As explained in \cite{Hull4}, for any of the following conjugacy class $\mathcal{M}$, $-\mathcal{M}$ and $\pm \mathcal{M}^{-1}$ are also conjugacy class. In the following we review the different cases:
\begin{enumerate}
    \item Parabolic case: There are an infinite number of parabolic $SL(2,\mathbb{Z})$ conjugacy classes with $Tr(\mathcal{M})=2$
    \begin{eqnarray}
        \mathcal{M}_p = \left( \begin{array}{cc}
           1  & n \\
            0 & 1
        \end{array}\right) \label{parabolicmonodromyUV}
    \end{eqnarray}
with $n\in \mathbb{Z}$. The real parameter of the parabolic supergravity is quantized in the higher dimensional theory.
\item Elliptic case: There are three elliptic $SL(2,\mathbb{Z})$ conjugacy classes with $\abs{Tr(\mathcal{M})<2}$ 
\begin{eqnarray}
     \mathcal{M}_{Z_3} = \left( \begin{array}{cc}
           1  & 1 \\
            -1 & 0
        \end{array}\right), \,      \mathcal{M}_{Z_4} = \left( \begin{array}{cc}
           0  & 1 \\
            -1 & 0
        \end{array}\right), \,      \mathcal{M}_{Z_6} = \left( \begin{array}{cc}
           0  & 1 \\
            -1 & -1
        \end{array}\right), \label{ellipticmonodromyUV}
\end{eqnarray}
It usually goes unnoticed that these matrices are conjugate to particular rotations of the $SO(2)$ group. Indeed these monodromies correspond to $SO(2)$ gaugings at specific angles. Indeed,
\begin{eqnarray}
 \mathcal{M}_{Z_3} = U\mathcal{M}_e(\frac{\pi}{3})U^{-1} \, , \, \hspace{0.3cm} \mathcal{M}_{Z_4} = \mathcal{M}_e(\frac{\pi}{2})\, , \, \hspace{0.3cm} \mathcal{M}_{Z_6} = V\mathcal{M}_e(\frac{2\pi}{3})V^{-1}
\end{eqnarray}   
Therefore, the compact parameter of the $SO(2)$ must be restricted to discrete values. 
\item Hyperbolic case: There are an infinite number of hyperbolic conjugacy classes with $\abs{Tr(\mathcal{M})}>2$ for $\abs{n}\ge 3$, generated by
\begin{eqnarray}
\mathcal{M}_{h} = \left( \begin{array}{cc}
           n  & 1 \\
            -1 & 0
        \end{array}\right) \label{hiperbolicmonodromyUV}
\end{eqnarray}
and it can be seen they are conjugate to $\mathcal{M}_h$ given by (\ref{hiperbolicmonodromy}). It is not difficult to see that it requires the parameter $a$ to be expressed in terms of n as
\begin{eqnarray}
    a=\ln{\left( \frac{n}{2}+\sqrt{\frac{n^2}{4}-4}\right)}
\end{eqnarray} 
such that $n$ is discrete. Once again we can observe that the the parameter $a$ become restricted to be particular integers.
\end{enumerate}

Besides the seven massive deformations of type II supergravities in 9d from figure (\ref{fig1}), one may also consider combining mass parameters in such way that is consistent with supersymmetry. Such cases were studied in \cite{Bergshoeff9,Bergshoeff5} and will not be considered in this work. There are also several domain wall solutions that has been obtained and expressed in terms of branes of type IIB string theory \cite{Bergshoeff9,Cowdall}. 

In this work, we will go beyond the work done in \cite{mpgm2} and provide a more complete quantum description of type IIB gauged supergravities in 9d in terms of the M2-brane with nontrivial fluxes on a torus. The corresponding Hamiltonian has a $SU(N)$ regularized model with a discrete supersymmetric spectrum. The global description is given in terms of twisted torus bundles with monodromy contained in $SL(2,Z)$ \cite {mpgm10}. The group transformation between equivalent twisted torus bundles for a given flux and monodromy seems to correspond with the restricted gauge symmetries at low energies. This allows us to give a $D=11$ interpretation to the discrete values that the mass parameters shall take in the quantum description for each case. The global symmetry will correspond to the quantization of $SL(2,\mathbb{R})$ and it will be, in general, a discrete subgroup contained in $SL(2,\mathbb{Z})$. 

\section{Local and global description of quantum M2-branes with fluxes and monodromy}
In this section, we will formulate the worldvolume description of M2-branes with nontrivial winding on a torus. This theory has a $SU(N)$ regularized model with a discrete supersymmetric spectrum. The global description is given in terms of equivalence classes of twisted torus bundles with monodromy contained in $SL(2,\mathbb{Z})$.

\subsection{M2-branes with nonzero winding on a torus.}

The light cone gauge (LCG) bosonic Hamiltonian for an M2-brane in the presence of a non-vanishing three-form background was given in \cite{deWit}, where the authors have considered a general spacetime with metric $G_{\mu\nu}$ written in a convenient form using the gauge $G_{--}=G_{a-}=0$. Its supersymmetric extension on a local Minkowski spacetime ($M_{11}$) was obtained in \cite{mpgm6} (where it was also shown its consistency with $C_{\mu\nu\rho}$ constant, not necessarily zero) and it is given by 
\begin{eqnarray}\label{HCM2}
\mathcal{H}&=&\left[\frac{1}{(\widehat{P}_--TC_-)}\left(\frac{1}{2}(\widehat{P}_a-TC_a)^2+\frac{T^2}{4}(\epsilon^{uv}\partial_u X^a \partial_v X^b)^2\right) \right. \nonumber  \\
&-& \left. T\bar{\theta}\Gamma^-\Gamma_a \left\lbrace X^a,\theta \right\rbrace - TC_{+-}- TC_+ \right], 
\end{eqnarray}
subject to the first and second-class constraints
\begin{eqnarray}
\widehat{P}_a\partial_u X^a + \widehat{P}_- \partial_u X^- + \bar{S}\partial_u \theta &\approx& 0, \\
S - (\widehat{P}_--TC_-)\Gamma^- \theta &\approx& 0 ,
\end{eqnarray}
with $T$ being the M2-brane tension and the unique free parameter of the theory, $\widehat{P}_a$ the canonical conjugate to $X^a$ and $S,\bar{S}$ are the conjugate momenta to $\bar{\theta},\theta$ (Majorana spinors in 11D), respectively \cite{deWit2}. 

The embedding used in this paper is the same one used in \cite{deWit2,deWit6} seminal papers. We are considering the light cone gauge and hence the space-time indices $\mu,\nu,\rho=0,\dots,10$ are split according to $\mu=(+,-, a)$, where $a=1,\dots,9$ are the transverse indices to the null light coordinates (see for example \cite{deWit2}). The worldvolume indices are $i=0,1,2$, with $u,v=1,2$ labeling the spatial coordinates. We are considering an embedding of the M2-brane on the complete 11D space-time. That is, $X^a(\sigma^1,\sigma^2,\tau)$ are maps from $\Sigma$, a Riemann surface of genus 1, to the target space, $X^a: \Sigma \rightarrow M_{11}$.

The LCG three-form components are written according to \cite{deWit} as
\begin{equation}\label{CaLCG}
\small
\begin{aligned}
& C_a  =  -\epsilon^{uv}\partial_uX^- \partial_vX^b C_{-ab} +\frac{1}{2}\epsilon^{uv}\partial_uX^b \partial_vX^c C_{abc} \, , \\
& C_{\pm}  =  \frac{1}{2}\epsilon^{uv}\partial_uX^a \partial_vX^b C_{\pm ab} \,, \qquad C_{+-}  =  \epsilon^{uv}\partial_uX^- \partial_vX^a C_{+-a} \,,
\end{aligned}
\end{equation}
where $C_{+-a}=0$ is fixed by gauge invariance of the three-form and $C_{\pm ab}$ and $C_{abc}$ are assumed, in this work, to be nontrivial constants by background fixing. Let us notice that $X^-$ appears explicitly in the Hamiltonian through $C_a$ \cite{deWit}. Nevertheless, one may perform a canonical transformation of the Hamiltonian by performing the following change of variables \cite{mpgm6}
\begin{eqnarray}
   P_a=\widehat{P}_a-TC_a, \quad P_-=\widehat{P}_--TC_-, \quad S = \widehat{S}, \quad \widehat{X}^a=X^a, \quad \widehat{X}^-=X^-, \quad \widehat{\theta}=\theta. \nonumber 
\end{eqnarray}
 We may use the residual gauge symmetry generated by the constraints to impose the gauge fixing condition $P_-=P^0_-\sqrt{w}$, with $\sqrt{w}$ a regular density on the worldvolume. We can then eliminate ($X^-,P_-^0$) as canonical variables and obtain a formulation solely in terms of ($X^a,P_a$) and ($\theta,\bar{S}$).

If we consider a compactification of the target space, on a flat torus $T^2$ characterized by the Teichmuller parameter $\tau\in\mathbb{C}$ with $\mbox{Im}(\tau)> 0$ and a radius $R\in\mathbb{R}$, the embedding maps are splitted into the noncompact and compact sectors as follows
\begin{eqnarray}
X^a(\sigma^1,\sigma^2,\tau)=(X^m(\sigma^1,\sigma^2,\tau),X^r(\sigma^1,\sigma^2,\tau)) , \label{Embedding}
\end{eqnarray}
with $m=1,\dots,7$ and $r=8,9$, respectively. We may perform a Hodge decomposition on the closed, but not exact, one-forms $dX^r=dX_h^r + dA^r$, where $dX_h^r$ are the harmonic one-forms and $dA^r$ are the exact ones. $dX_h^r$ may be written in terms of a normalized basis of harmonic one-forms $d\hat{X}^r$ as $dX_h^1+idX_h^2 = 2\pi R(l_r+m_r\tau)d\hat{X}^r$. The wrapping condition on the compact sector is given by
\begin{equation}
\label{ec2notas}
\oint_{\mathcal{C}_r} d \left(X^8 + iX^9 \right)= 2 \pi R \left(l_r + m_r \tau \right) \, \in  \mathcal{L} \,,
\end{equation}
where $\mathcal{C}_r$ denotes the homology basis on $\Sigma$, $\mathcal{L}$ is a lattice on the complex plane ($\mathbb{C}$) such that $T^2=\mathbb{C}/\mathcal{L}$ and the winding numbers $l_r,m_r$ defines the wrapping matrix
 \begin{eqnarray}
     \mathbb{W} = \left(\begin{array}{cc}
         l_8 & l_9 \\
          m_8 & m_9
     \end{array} \right). \label{windingmatrix}
 \end{eqnarray}
So far, the Hamiltonian of the M2-brane on a torus can be written as
\begin{eqnarray}
H^{T^2}&=&\frac{1}{2P^0_-}\int_\Sigma d^2\sigma \sqrt{w}\left[\Big(\frac{P_m}{\sqrt{w}}\Big)^2+\Big(\frac{P_r}{\sqrt{w}}\Big)^2 + \frac{T^2}{2}\left(\left\{X^m,X^n\right\}^2 + 2\left\{X^m,X^r\right\}^2 \right. \right. \nonumber \\
&+&\left. \left. \left\{X^r,X^s\right\}^2\right)\right]- \frac{T}{2P^0_-}\int_\Sigma d^2\sigma \sqrt{w} (\bar{\theta}\Gamma_-\Gamma_r\left\{X^r,\theta\right\}-T\bar{\theta}\Gamma_-\Gamma_m\left\{X^m,\theta\right\}),\nonumber \label{HamiltonianM2}
\end{eqnarray}
where $\displaystyle \left\lbrace \bullet, \bullet \right\rbrace = \frac{\epsilon^{uv}}{\sqrt{w}}\partial_u \bullet \partial_v \bullet$.
Let us make some comments about this Hamiltonian. It can be seen that the harmonic maps on the compact sector may degenerate. Moreover, the string spikes identified of the M2-brane in $M_{11}$ and in $M_{10}\times S^1$, are also present in this torus compactification. Consequently, it has a continuous spectra \cite{deWit6,deWit3,deWit4}. These two matters were addressed during the last decades (see \cite{Restuccia,Restuccia3,Boulton,mpgm6}) and will be resumed in the following paragraphs by ensuring the nonvanishing feature of the determinant of the wrapping matrix.

Once the dependence on $X^-$ has been eliminated, a quantization condition on $C_{\pm}$ can be imposed. This condition corresponds to a 2-form flux condition on the target space 2-torus, whose pull-back through $X_h^r$, with $r=8,9$, generates a 2-form flux condition on the M2-brane worldvolume as follows \cite{mpgm10}
 \begin{eqnarray}\label{fluxpullback}
 \int_{T^2}C_{\pm}=\frac{1}{2} \int_{T^2}C_{\pm rs} d\widetilde{X}^r\wedge d\widetilde{X}^s = \frac{1}{2} \int_{T^2}C_{\pm rs} d\widehat{X}^r\wedge d\widehat{X}^s =c_{\pm}\int_\Sigma \widehat{F} = k_{\pm},
 \end{eqnarray}
 where $C_{\pm rs}=c_{\pm}\epsilon_{rs}$ with $c_{\pm}\in \mathbb{Z}/\{0\}$, $\widetilde{X}^r$ are local coordinates on $T^2$, $k_{\pm}=nc_{\pm}$ with $n\in \mathbb{Z}/\{0\}$ and  $\widehat{F}$ is a closed 2-form defined on $\Sigma$ such that it describes a worldvolume flux condition
 \begin{equation}\label{central charge}
   \int_{\Sigma}\widehat{F} = \frac{1}{2}\int_{\Sigma}\epsilon_{rs}d\widehat{X}^r \wedge d\widehat{X}^s =n,
  \end{equation}
  where the integer $n=det(\mathbb{W})\ne 0$ characterizing the irreducibility of the wrapping, where $\mathbb{W}$ is the winding matrix. Consequently, $C_{\pm}$ is a closed two-form defined on the target space torus. Indeed, the flux condition on $T^2$ implies a flux condition on $\Sigma$ which is known as 'central charge condition'. The irreducible wrapping condition ensures that the harmonic modes are nontrivial and independent.

The Hamiltonian of the M2-brane with $C_-$ fluxes becomes
\begin{eqnarray}
H^{C_-}&=&\frac{1}{2P^0_-}\int_\Sigma d^2\sigma \sqrt{w}\left[\Big(\frac{P_m}{\sqrt{w}}\Big)^2+\Big(\frac{P_r}{\sqrt{w}}\Big)^2 + \frac{T^2}{2}\left(\left\{X^m,X^n\right\}^2 + 2(\mathcal{D}_rX^m)^2 \right. \right. \nonumber \\
&+&\left. \left. (\mathcal{F}_{rs})^2+ (\widehat{F}_{rs})^2\right)\right]- \frac{T}{2P^0_-}\int_\Sigma d^2\sigma \sqrt{w} (\bar{\theta}\Gamma_-\Gamma_r\mathcal{D}_r\theta-T\bar{\theta}\Gamma_-\Gamma_m\left\{X^m,\theta\right\}),\nonumber \label{HamiltonianM2}
\end{eqnarray}
and 
\begin{eqnarray}\label{HamiltonianC+}
     H^{C_+} &=& H^{C_-} - 2\widehat{P}_-^0 T  \int d^2\sigma \sqrt{w}C_+,
\end{eqnarray}
only differs in a constant term \cite{mpgm6}. Interestingly,  the supermembrane on $M_9\times T^2$ with a central charge condition associated with an irreducible wrapping \cite{Restuccia}, is equivalent to the Hamiltonian of a supermembrane on $M_9^{LCG}\times T^2$ on a quantized $C_-$ background, i.e. $\mathcal{H}^{CC}=\mathcal{H}^{C_-}$. Hence, the discreteness property of the former automatically implies the discreteness of the M2-brane with $C_{\pm}$ fluxes. When $C_+\ne 0$, the spectrum is discrete and shifted by a constant value.

The degrees of freedom of the theory are $X^m,A^r,\theta$. Let us emphasize that, because of the general embedding considered in (\ref{Embedding}), these M2-branes are not, in general, free particles in the noncompact space, but extended membranes -toroidal for the case considered here- that contain a non-lineal quartic bosonic potential and its fermionic counterpart.

On the other hand, the symplectic covariant derivative is defined as \cite{Ovalle1}
\begin{eqnarray}
   \mathcal{D}_rX^m &=&D_rX^m+\left\{ A_r,X^m\right\}, \label{symp-cov-der}
\end{eqnarray}
with $D_r$ is a covariant derivative defined as \cite{mpgm2,mpgm7} and it satisfies 
\begin{eqnarray}
    (D_8+iD_9) \, \bullet  = 2\pi R (l_r+m_r\tau)\left\lbrace \widehat{X}^r,\, \bullet \right\rbrace , \nonumber 
\end{eqnarray}
 The gauge contribution is given by 
 $\widehat{F}$ the minimal curvature related to the worldvolume flux on $\Sigma$ (\ref{fluxpullback}) and 
\begin{eqnarray}
  \mathcal{F}_{rs}&=& D_rA_s-D_sA_r+\left\{ A_r,A_s\right\}, \label{Fsymp}
\end{eqnarray}
corresponds to a symplectic curvature associated to the one-form connection $A_r dX^r$, where $A^r$ contains the dynamical degrees of freedom related to the exact sector of the map on $T^2$.

This Hamiltonian is subject to the local and global constraints associated to the area preserving diffeomorphisms (APD) 
\begin{eqnarray}
\small \left\{ \frac{P_m}{\sqrt{w}} , X^m\right\} + \mathcal{D}_r\left( \frac{P_r}{\sqrt{w}}\right)+\left\lbrace \frac{\bar{S}}{\sqrt{w}},\theta \right\rbrace  &\approx& 0, \label{LocalAPD}\\
 \oint_{C_S}\left[\frac{P_m dX^m}{\sqrt{w}} + \frac{P_r (dX_h^r+dA^r)}{\sqrt{w}} + \frac{\bar{S} d\theta}{\sqrt{w}}\right] &\approx& 0, \label{GlobalAPD}
\end{eqnarray}
which appear as a residual symmetry on the theory after imposing the LCG in the covariant formulation. In fact, we have shown that M2-branes with $C_{\pm}$ fluxes are invariant under the full group of simplectomorphisms, which considers the sectors connected and not connected to the identity. Furthermore, symplectomorphisms on $T^2$ are in one-to-one correspondence to symplectomorphisms on $\Sigma$  \cite{mpgm10}. 

Classically, this Hamiltonian does not contain string-like spikes at zero cost energy that may produce instabilites  \cite{mpgm}. At quantum level the $SU(N)$ regularized theory has a purely discrete spectrum since it satisfy the sufficiency criteria for discreteness found in \cite{Boulton}. The theory preserves $1/2$ when considering a minimal configuration of KK or winding charges, and $1/4$ when considering a more general state with both, winding and KK \cite{mpgm6} on the interior of the Moduli space. 

\subsection{M2-branes on twisted torus bundles with monodromy}
The M2-branes with $C_\pm$ fluxes can be formulated on twisted torus bundles with monodromy in $SL(2,Z)$ \cite{mpgm10}. In fact, the $U(1)$ principle bundle associated with the nontrivial quantized fluxes, or to the central charge condition, is compatible with the formulation of the M2-brane on a symplectic torus bundle, with structure group, the symplectomorphisms preserving the $U(1)$ curvature. There exists a natural homomorphism
\begin{eqnarray}\label{monodromy}
\mathcal{M}_G:\Pi_1(\Sigma)\rightarrow \Pi_0(Symp(T^2))=SL(2,Z).
\end{eqnarray}
The subgroup of $SL(2,Z)$ determined by the homomorphism is called the monodromy of the formulation. The classification of symplectic torus bundles with monodromy in terms of $H^2(\Sigma, \mathbb{Z}^2_{\rho})$ was found by \cite{Kahn}. In the aforementioned paper it is shown the existence of a one-to-one correspondence between the inequivalent classes of symplectic torus bundles for a given monodromy conjugacy class inducing the module structure $Z_\rho^2$ on $H_1(T^2)$ and the elements of $H^2(\Sigma,Z_\rho^2)$, the second cohomology group of the bundle with base $\Sigma$ and coefficients in $Z_\rho^2$. This homomorphism gives to each homology and coholomogy group on the bundle the structure of $Z\left[\Pi_1(\Sigma)\right]$-module.  It classifies the symplectic torus bundles for a given monodromy in terms of the characteristic class. Hence, the symplectic torus bundles, with base manifold a torus, are
classified, for a given monodromy, according to the inequivalent coinvariants \cite{mpgm7,mpgm2}.

Therefore, sectors of M2-branes on $M_9\times T^2$ with the irreducible wrapping condition, contain two compatible gauge structures. The first one is given by the symplectic structure of the bundle, which ensures the existence of a symplectic connection under symplectomorphisms. The second gauge structure is a nontrivial $U(1)$ principal bundle related to the 2-form flux on $\Sigma$ due to the central charge condition or the 2-form flux condition on the target-space.

In \cite{mpgm10} it was proved that the symplectic structure and the U(1) principal bundle are related and generate a twisted torus bundle,
\begin{equation}
\label{ec7notasseccion6.2}
\mathbb{T}_W^3\equiv {T}_{U(1)}^2 \rightarrow E' \rightarrow \Sigma \, ,  
\end{equation}
where the base manifold is given by the worldvolume Riemann surface $\Sigma$, the fiber is a twisted torus $\mathbb{T}^3$, given by the U(1) principal bundle associated with the nontrivial flux condition on $T^2$.

\subsubsection{Trivial monodromy}
In \cite{mpgm3} two inequivalent $SL(2,Z)$ symmetries of the M2-brane with central charges were identified. One is associated with the target torus and will be denoted as $SL(2,Z)_{T^2}$, while the other is associated with the base manifold and will be denoted as $SL(2,Z)_{\Sigma}$. In \cite{mpgm10} M2-branes  with $C_{\pm}$ fluxes were shown to be  invariant under the 2-dimensional area preserving diffeomorphisms, or equivalently, 2-dimensional symplectomorphisms, connected and not connected to the identity. The invariance of the Hamiltonian under those connected with the identity is guaranteed by the first class constraint of the theory. The isotopy classes of symplectomorphisms on the base manifold determine a group, which in this case is $SL(2,Z)_{\Sigma}$. The symplectomorphisms not connected to the identity change the homology basis on $\Sigma$ together with the corresponding basis of harmonic one-forms and the winding matrix as follows
	\begin{eqnarray}
	    d\widetilde{X}^r \rightarrow (S_1^*)^r_sd\widehat{X}^s \label{SL(2,Z)sigma1},\quad
	    \mathbb{W} \rightarrow \mathbb{W}(S_1^*)^{-1} , \label{SL(2,Z)sigma2}
	\end{eqnarray}
	with $S_1^*\in SL(2,Z)_\Sigma$.
The symplectomorphisms not connected with the identity on the target $T^2$ are the ones that change the moduli of the 2-torus by a modular transformation \cite{mpgm23} as follows,
\begin{eqnarray}
	\tau&\rightarrow& \tau'=\frac{a\tau+b}{c\tau+d} , \quad
	R\rightarrow R'= R|c\tau+d| , \quad
	A\rightarrow A'= Ae^{i\varphi_\tau}, \nonumber \\	
	\mathbb{W}&\rightarrow&\mathbb{W}'=S_2^*\mathbb{W},  \quad
	Q\rightarrow Q'=S_2Q, \quad \Gamma \rightarrow \Gamma'=\Gamma e^{i\varphi_\tau}	\label{SdualQ}
	\end{eqnarray}
with $S_2$, $S_2^*$ matrices of $SL(2,Z)_{T^2}$ given by
	\begin{eqnarray}
	S_2 = \left(\begin{array}{cc}
	  a   & b \\
	    c & d
	\end{array}\right), \quad
	S_2^* = \left(\begin{array}{cc}
	  a   & -b \\
	    -c & d
	\end{array}\right), \quad
	    c\tau+d = |c\tau+d|e^{-i\varphi_\tau} . \label{Sdualphi} 
	\end{eqnarray}
and $\Gamma=\Gamma_8+i\Gamma_9$ is the complex gamma matrix present in the fermionic term and related to the compact directions. The full supersymmetric Hamiltonian is  invariant under (\ref{SdualQ}) \cite{mpgm23}.

The irreducible wrapping condition ensures a one-to-one correspondence of symplectomorphisms on $T^2$ and $\Sigma$, also assumed to be a 2-torus.

 Finally, the full M2-brane with $C_\pm$ fluxes mass operator it was given in \cite{mpgm22,mpgm23}. The KK and winding term were first obtained in \cite{Schwarz6} and in \cite{mpgm23} it is shown that they are strictly related to the M2-brane with a central charge condition associated with the irreducibility of the wrapping or with  the presence $C_\pm$ fluxes  on $M_9\times T^2$. Double dimensional reduction of this mass operator reproduce the full mass operator of the type IIB $SL(2,\mathbb{Z})$ ($p,q$)-string winding on a circle as it is shown in \cite{mpgm23}.

\subsubsection{Nontrivial monodromy}
A symplectic torus bundle is defined by $E$ the total space, $F$ the fiber which is the torus of the target-space $T^2$ compact sector and the base space $\Sigma$, which is also a torus. The structure group $G$ corresponds to the group of the symplectomorphism preserving the canonical symplectic two-form on $T^2$. On $\Sigma$, there exists an induced symplectic two-form, obtained from the pullback of the two-form on $T^2$ by the harmonic map from $\Sigma$ to the fiber $T^2$. We notice that the group of symplectomorphisms in $T^2$ or $\Sigma$ is isomorphic to the area preserving diffeomorphisms. The symplectomorphisms in $T^2$ and in $\Sigma$ define the isotopic classes with a group structure $\Pi_0(G)$, in the case under consideration $SL(2,Z)$.

The action of $G$ on the fiber produces an action on the homology and cohomology classes of $T^2$. It reduces to an action of $\Pi_0(G)$. Besides, there is an homomorphism (\ref{monodromy}). Each homomorphism defines a linear representation 
\begin{eqnarray}
     \rho:\Pi_1(\Sigma)\rightarrow SL(2,Z),
\end{eqnarray}
acting on the first homology group in $T^2$, $H_1(T^2)$. Because $H_1(T^2)$ is an abelian group, this homomorphism gives the structure of the $Z(\left[\Pi_1(\Sigma) \right])$-module to each homology and cohomology group on the bundle. Given a monodromy, \cite{Kahn} established the existence of a one-to-one correspondence between the equivalent classes of symplectic torus bundles, induced by the module structure $Z_\rho^2$ on $H_1(T^2)$, and the elements of $H^2(\Sigma, Z_\rho^2)$, the second cohomology group of $\Sigma$ with coefficients $Z_{\rho}^2$. They classify the symplectic torus bundles for a given monodromy in terms of the characteristic class. In the case of a symplectic torus bundle  with base a torus $\Sigma$, the classification in terms of these characteristic classes is equivalent to the classification in terms of coinvariant classes of the monodromy subgroup, acting on ($p,q$) charges. We denote the coinvariant classes simply as coinvariants. As we have previously explained, the M2-branes on $M_9\times T^2$ with worldvolume flux -due either to an irreducible wrapping condition, or to an induced background flux- contain two compatible gauge structures that lead to the twisted torus bundle with monodromy

Let us consider that the monodromy on the fiber is given by
\begin{eqnarray}\label{monodromyfiber}
    \mathcal{M}_G=\begin{pmatrix} \mathcal{M}_{11} & \mathcal{M}_{12}\\
                        \mathcal{M}_{21} & \mathcal{M}_{22}
 \end{pmatrix}^{(\alpha+\beta)} \in SL(2,Z),
\end{eqnarray}
where ($\alpha,\beta$) are the integers characterizing the elements of $\Pi_1(\Sigma)$ and specific values of $\mathcal{M}_{ij}$, with $i,j=1,2$ will lead to parabolic, elliptic or hyperbolic monodromies according to its trace. The induced transformation on $\Sigma$, also called induced monodromy on $\Sigma$, is given by
\begin{eqnarray}
   \mathcal{M}_G^*= \Omega^{-1}\mathcal{M}_G(\alpha,\beta)\Omega = \begin{pmatrix} \mathcal{M}_{11} & -\mathcal{M}_{12}\\
                        -\mathcal{M}_{21} & \mathcal{M}_{22}
 \end{pmatrix}^{(\alpha+\beta)} ,\label{inducedmonodromy}
\end{eqnarray}
with $\displaystyle \Omega=\left(\begin{array}{cc}
    -1 & 0  \\
    0 & 1
\end{array}\right)$, equivalently to $S^*_2$ in (\ref{SdualQ}). Therefore symplectomorphisms not connected with the identity on $\Sigma$ are realized by
	\begin{eqnarray}
	  d\widetilde{X}^r \rightarrow (\tilde{g}^*)^r_s d\widehat{X}^s \label{SL(2,Z)sigma11}, \quad
	    \mathbb{W} \rightarrow \mathbb{W}(\tilde{g}^*)^{-1}, \label{SL(2,Z)sigma22} 
	\end{eqnarray}
where $\tilde{g}^*\in\mathcal{M}^*_G$ and the action of $S_U$-duality, when the monodromy is nontrivial, is given by
\begin{eqnarray}
	\tau &\rightarrow& \frac{a\tau+b}{c\tau+d} , \label{GSdualtau}
\quad	R\rightarrow R|c\tau+d| , \label{GSdualR} \quad
	A\rightarrow Ae^{i\varphi_\tau},  \quad
	 \\
	\Gamma &\rightarrow& \Gamma e^{i\varphi_\tau}, \quad \mathbb{W}\rightarrow g^*\mathbb{W},  \label{GSdualW}\quad
	Q \rightarrow gQ,	\label{GSdualQ}
	\end{eqnarray}
	with $\displaystyle g=\left( \begin{array}{cc}
	 a  & b  \\
	    c & d
	\end{array} \right) \in\mathcal{M}_G$ and $c\tau+d = |c\tau+d|e^{-i\varphi_\tau}$. The M2-brane sectors with central charges are invariant under these $SL(2,Z)$ symmetry transformations on $\Sigma$ and $T^2$.

The winding and KK term on the mass operator of the M2-brane with fluxes and monodromy can be computed and they give
\begin{eqnarray}
    (TnA_{T^2})^2 + b^2 \frac{\vert q\tau-p \vert^2 }{(R\mbox{Im}(\tau))^2}
\end{eqnarray}
where $n$ denotes the central charge units, $p$, $q$ are the KK charges and $b$ is a constant of units $(energy)\times(length)$.

 So far, the mass operator of supermembranes with monodromy contained in $SL(2,Z)$ can be written as
 \begin{eqnarray}
M_{C_\pm}^2 &=&  (TnA_{T^2})^2 + b^2 m^2 \vert \tau_R^T Q\vert^2 +2\widehat{P}_-^0H'^{C_\pm},\label{MassOp_monodromy}
 \end{eqnarray}
with $H'^{C_\pm}$ the nonzero modes of the Hamiltonian of the M2-brane with $C_\pm$ fluxes and nontrivial monodromy and
\begin{eqnarray}
    \tau_R^T = \frac{1}{R\mbox{Im}(\tau)}\left( \begin{array}{cc}
        -1 ,& \tau 
    \end{array}\right), \quad Q = \left( \begin{array}{c}
         p  \\
         q 
    \end{array} \right)
\end{eqnarray}
\section{M2-brane on the module of $\mathcal{M}_g$ coinvariants}
In this section, we present our new results. We show that the worldvolume description of the M2-brane with nontrivial winding on a torus, i.e. nontrivial worldvolume fluxes, which has a discrete supersymmetric spectrum, reproduces the type II gauged supergravities in $D=9$ at low energies. We claim that the origin of the gauged symmetry group is related with the equivalence class of twisted torus bundle for a given monodromy. 

Let us consider the supersymmetric mass operator (\ref{MassOp_monodromy}) corresponding to the M2-branes with fluxes and nontrivial monodromy contained in $SL(2,\mathbb{Z})$. In \cite{mpgm23} we restricted ourselves to parabolic monodromies while in this paper we consider all conjugacy classes. 

Let us emphasize that the contribution of the monodromy is nontrivial. The mass operator in \cite{Schwarz6} was obtained for M2-branes on a torus with trivial monodromy, in this case each pair of charges ($p,q$) determines an equivalence class related among them by the $SL(2,Z)$ symmetry. The main point is that when the monodromy is not trivial, there are equivalence classes that contains a set of ($p,q$) charges related by an internal symmetry. These sets are denoted as coinvariant classes. Furthermore, there is also a symmetry relating the different classes among them. In what follows we will determine both symmetries. 

The mass operator of the M2-brane with nonvanishing winding on a torus depends on the KK charges but also on the moduli and winding number. In particular, one has to give the associated moduli to the corresponding coinvariant. In the trivial monodromy case one has to provide the moduli for each pair ($p,q$). For nontrivial monodromies the internal symmetry define an equivalence class of charges $(p,q)$ and moduli parameters which leave invariant the coinvariant together with the mass operator. In this way, the theory is formulated in terms of equivalence classes. This is reminiscent of what occurs with the gauge theories, which are defined in terms of equivalence classes, elements on the same class are related by a gauge symmetry. In this sense, we argue that this internal discrete symmetry is the origin of the gauge symmetry in gauged supergravity.

\subsection{The symplectic connection as the gauge vector of type II supergravity }

The twisted torus bundles are an evidence of the compatibility of two different gauge structures (nontrivial $U(1)$ and symplectic) on the global description of the M2-branes with nontrivial winding. The transformation of the symplectic connection under symplectomorphisms (connected and not connected with the identity) is given by
\begin{eqnarray}\label{SympconnecM2}
    \delta A^r = -\mathcal{D}_r\xi
\end{eqnarray}
when $\delta X_h^r=0$, \cite{mpgm10}. The symplectic covariant derivative is given by (\ref{symp-cov-der}) and $\xi$ is the parameter of the transformation. 

On the other hand, the gauge vector of the type II supergravities in nine dimensions are identified with the one of the three one-forms $A^I$ with $I=0,1,2$ \cite{Bergshoeff5}. From figure \ref{fig1}, it can be seen that $A^0$ is the graviphoton and it corresponds to the gauge vector on the type IIB sector and on the KK reduction of the type IIA Romans supergravity to 9d. Similarly, $A^1$, $A^2$ which comes from the RR and NSNS two forms, respectively, are the gauge vectors acting on  the rest of type IIA gauged supergravities.

We have already mentioned that the type II gauge supergravities may have a higher dimensional origin via Scherk Schwarz reduction or they can be produced by a deformation through the embedding tensor formalism. The transformation of the background fields is the same in both approaches. In terms of the embedding tensor, the transformation of the one-forms $A^I$ can be written as \cite{Melgarejo}
\begin{eqnarray}
    \delta_\Lambda A^I = -\mathcal{D}\Lambda^I + Z^I_i\Lambda^i, \label{gaugevector_trans}
\end{eqnarray}
where $\Lambda$ is the parameter of the transformation and
\begin{eqnarray}
    Z^0_i &=& 3\Theta^4_i + \frac{1}{2}\Theta^5_i, \\
    Z^i_j &=& \Theta^n_0 (T_n)^i_j - \frac{3}{4}\Theta_0^5\delta^1_j\delta^i_1.
\end{eqnarray}
with $\Theta_i^n$, $n=1,2,3$ the embedding tensor and $T_n$ the three generators of $SL(2,\mathbb{R})$.

For the four type IIA gauge supergravities in nine dimensions, the components of the embedding tensor are given by \cite{Melgarejo}
\begin{eqnarray}
    \Theta_1^4=-m_{11} \, , \hspace{0.3cm} \Theta_2^4=m_{IIA} \, , \hspace{0.3cm}
    \Theta_1^5=\widetilde{m}_4 \, , \hspace{0.3cm}
    \Theta_2^5=m_{4} \, , \hspace{0.3cm}
\end{eqnarray}
while for the type IIB sector
\begin{eqnarray}
    \Theta^m_0 &=& m_n \, , \hspace{0.3cm} \Theta_0^5 = -\frac{16}{3} m_{IIB}
\end{eqnarray}
where $n=1,2,3$ corresponds to the triplet according to the previous section.

It can be seen from (\ref{gaugevector_trans}) that the gauge vector for all type II gauge supergravities transform as 
\begin{eqnarray}
    \delta A^I = -\mathcal{D}\Lambda^I
\end{eqnarray}
where $I=1,2,3$ corresponds to the three one-forms of type II supergravity in nine dimensions. This transformation is reminiscent from the transformation of the symplectic connection of the M2-branes with fluxes and monodromy (\ref{SympconnecM2}). The symplectic gauge structure allows us to identify several equivalence classes of twisted torus bundles. We will try to dive into this relation in the next section.

\subsection{Equivalence classes of twisted torus bundles with monodromies in $SL(2,Z)$}
 Inequivalent torus bundles are classified according to the coinvariant classes, briefly coinvariants, for a given monodromy \cite{mpgm3,mpgm2,mpgm7,mpgm10}. The coinvariants related to the fiber and the base manifold are given by 
 \begin{eqnarray}
      C_F &=& \left\lbrace Q + \mathcal{M}_g\widehat{Q}-\widehat{Q}\right\rbrace, \label{C_F}\\
      C_B &=& \left\lbrace W + \mathcal{M}_g^*\widehat{W}-\widehat{W}\right\rbrace, \label{CB}
 \end{eqnarray}
respectively, where $Q = \left( \begin{array}{c}
         p  \\
          q 
     \end{array}\right) $ with $p,q\in\mathbb{Z}$ and $W = \left( \begin{array}{c}
         l_1  \\
          m_1 
     \end{array}\right)$ with $l_1,m_1\in \mathbb{Z}$, $Q$ and $W$ are KK and winding charges, $\widehat{Q}$ and $\widehat{W}$ correspond to arbitrary charges and $\mathcal{M}_g$ is the monodromy subgroup (parabolic, elliptic or hyperbolic). Given $\mathbb{W}$ as in (\ref{windingmatrix}), we will consider the class of matrices given by 
     \begin{eqnarray}
    \mathbb{W} = \left(\begin{array}{cc}
        l_1  & l_2 \\
         m_1 & m_2
     \end{array} \right)\left(\begin{array}{cc}
        1  & \frac{\lambda}{m} \\
         0 & 1
     \end{array} \right)    
     \end{eqnarray}
     with $\lambda,m \in\mathbb{Z}$ and $l_1=ml_1'$ and $m_1=mm_1'$ with $l_1',m_1'$ relatively primes. These are the most general matrices with $W$ as first column and $det(\mathbb{W})=n$. We use the first column of $\mathbb{W}$ in the definition of $C_B$, but we could have used the second column also. The following reasoning is also valid in both cases. 

     As formerly discussed in \cite{mpgm7}, if the monodromy is trivial, the coinvariants contain only one element,  but for a nontrivial monodromy class, the coinvariants associated with the base and the fiber contain an equivalence class of KK and winding charges, respectively, related to the same bundle. As shown in \cite{mpgm2}, it is straightforward to see that the Hamiltonian of M2-branes with $C_\pm$ fluxes is invariant in an orbit of charges ($gQ\subset C_F$) generated by the monodromy, with $g\in\mathcal{M}_g$, restricting the  $SL(2,Z)_{T^2}$ transformation (\ref{GSdualtau})-(\ref{GSdualW}) by the corresponding monodromy subgroup. 

In \cite{mpgm23} we demonstrated that, for the case of the parabolic monodromy, the Hamiltonian is invariant not only in the associated orbit but in the complete  coinvariant. In this paper we extend this result to all the $SL(2,Z)$ monodromy subgroups analyzing their properties. In the following, the discussion will be limited to the coinvariants of the fiber associated with the KK charges. However, this can also be applied to the coinvariants of the base related to the windings. 

\paragraph{Parabolic subgroup:}
It is generated by the abelian parabolic subgroup
\begin{eqnarray}\label{parabolic_monodromy}
     \mathcal{M}_p = \left(\begin{array}{cc}
         1 & 1 \\
         0 & 1
     \end{array}\right) ^{(\alpha+\beta)}.
\end{eqnarray}
This parabolic representation contains the infinite inequivalent conjugate classes of parabolic monodromy
\begin{eqnarray}
   \mathcal{M}_p= \left( \begin{array}{cc}
        1 & k  \\
        0 & 1
    \end{array}\right), \label{parabolicrep}
\end{eqnarray}
with $k\in\mathbb{Z}$. The coinvariant of the fiber (\ref{C_F}) are given by
\begin{eqnarray}
     C_F &=& \left( \begin{array}{c}
         p +(\alpha+\beta)\widehat{q}  \\
          q 
     \end{array}\right) = \left( \begin{array}{c}
        \mathbb{Z}  \\
          q 
     \end{array}\right), \label{CFP}
\end{eqnarray}
 In this case, $C_F$ is characterized by $q$. The different values define inequivalent classes of torus bundles with parabolic monodromy. It can be seen that we have $\mathbb{Z}$ $q$-inequivalent coinvariants with $\mathbb{Z}$ element within the same coinvariant.

 \paragraph{Elliptic case:} There are three elliptic monodromy subgroups, isomorphic to the cyclic group $Z_k$, with $k=3,4,6$. These subgroups are generated by $\mathcal{M}_{Z_3}$, $\mathcal{M}_{Z_4}$ and $\mathcal{M}_{Z_6}$ given by (\ref{ellipticmonodromyUV}). The corresponding coinvariants are:
 \begin{eqnarray}
     C_{F(Z_k)} &=& \left( \begin{array}{c}
         p +\mathbb{Z}  \label{CFE}\\
          q +\mathbb{Z}-k\widehat{q}
     \end{array}\right)
\end{eqnarray}
It can be seen that, associated to each subgroup there are $k$ coinvariants, each one characterized by the elements $ \left( \begin{array}{c}
         0 \\
          0
     \end{array}\right)$ and $ \left( \begin{array}{c}
         0\\
          k'
     \end{array}\right)$, where $k'=1,\dots,k-1$.

     \paragraph{Hyperbolic case:} Hyperbolic monodromies are generated by matrices of the form (\ref{hiperbolicmonodromyUV}) with $\abs{n}\ge 3$ and their corresponding coinvariants can be written as
\begin{eqnarray}
     C_{F} &=& \left( \begin{array}{c}
         p +(n-2)\widehat{p}+\mathbb{Z}  \\
          q +\mathbb{Z}
     \end{array}\right), \label{CFH}
\end{eqnarray}
     It can be checked that there are $(n-2)$ coinvariants, each one characterized by the elements $ \left( \begin{array}{c}
         0 \\
          0
     \end{array}\right)$ and $ \left( \begin{array}{c}
         0\\
          k'
     \end{array}\right)$.
     
Therefore, we have that there are equivalence classes of KK charges for a given nontrivial monodromy contained in $SL(2,\mathbb{Z})$. It implies that the global description of the M2-brane with fluxes and monodromy is classified in the equivalence classes of twisted torus bundles. While in the parabolic and hyperbolic cases, there are infinite inequivalent bundles, the elliptic case is characterized by a finite number. Moreover, we can identify explicitly the elements of the coinvariants, which will be relevant to explain the allowed discrete values of the parameters in the low energy from the UV description. 

It is known that the mass operator given by (\ref{MassOp_monodromy}) is invariant under the monodromy $g\in \mathcal{M}_G$. Therefore, the Hamiltonian is consistently defined on the orbit of KK (winding) charges generated by the monodromy $g$ ($g^*$), which is contained in the coinvariant.

Nevertheless, it can be seen that the mass operator is invariant if
\begin{eqnarray}
     Q'&=& \Lambda Q, \label{transQ1} \\
     \mathbb{W}' &=& \Lambda^*\mathbb{W}, \label{transW1}\\
       \tau' &=& \frac{a\tau +b}{c\tau + d},\label{transtau} \\
       R' &=& R\abs{c\tau+d}, \label{transR}\\
       A &=& Ae^{-\varphi}, \\
       \Gamma &=& \Gamma e^{-i\varphi}
       \label{transtau1}
\end{eqnarray}
where the $\Lambda$ matrices given by
\begin{eqnarray}  
     \Lambda &=& \left( \begin{array}{cc}
         a & b  \\
         c & d 
     \end{array}\right) = \left( \begin{array}{cc}
        a  & \frac{1}{q}(p'-ap) \\
         \frac{1}{p'}(aq'-q) & \frac{1}{q}(q'-\frac{apq'}{p'} + \frac{qp}{p'})
     \end{array} \right),  \label{lambda_SL(2,Q)}\\
     \Lambda^* &=&\Omega^{-1}\Lambda \Omega
\end{eqnarray}
with $\displaystyle \Omega = \left( \begin{array}{cc}
    -1 & 0 \\
     0 & 1
\end{array}\right)$ and 
\begin{eqnarray}
    a = \frac{\pm Tr(\Lambda)qp'-q'p'-qp}{qp'-pq'}, \label{Aoflambda}
\end{eqnarray} 
and $p',q'$ are the elements of $Q'$. It is worth mentioning that $\Lambda$ is a matrix of $SL(2,\mathbb{R})$. However, they are not arbitrary matrices in the sense that they satisfy the quantization of charges (\ref{transQ1}). That is, integer KK charges $p,q$, are mapped to $p',q'$ integer charges. 

The transformation (\ref{transQ1}) maps every coinvariant for a given monodromy onto the same coinvariant. Therefore it maps any element of a given coinvariant onto the same coinvariant
\begin{eqnarray}\label{Paraboliccoinv}
     Q&\xrightarrow[]{\Lambda}& C_F,
\end{eqnarray}
where $\abs{Tr(\Lambda)}=2$ for parabolic monodromies, $\abs{Tr(\Lambda)}<2$ for elliptic monodromies and $\abs{Tr(\Lambda)}>2$ for hyperbolic monodromies. Consequently, the value of $a$ in (\ref{lambda_SL(2,Q)}) given by (\ref{Aoflambda}), will depend, not only on the initial and final elements, but on the monodromy subgroup.

Moreover, as it is shown in the appendix \ref{ApeA} the $\Lambda$ matrices for a given monodromy are conjugated to the one-parameter subgroups of $SL(2,\mathbb{R})$. Indeed, $\Lambda = U_g g U_g^{-1}$ where $U_g \in SL(2,\mathbb{R})$ and $g$ is a $SO(2)$ matrix for the elliptic case, an $SO(1,1)^+$ matrix for the hyperbolic case and a parabolic matrix for the parabolic with negative trace (for parabolic matrices with positive trace, $U_g=1$).

This transformation, together with the transformation of $\tau$ and $R$, leaves invariant KK term,
\begin{eqnarray}
     \frac{\vert q'\tau'-p'\vert^2 }{(R\mbox{Im}(\tau'))^2} = 
     \frac{\vert q\tau-p\vert^2 }{(R\mbox{Im}(\tau))^2}. 
\end{eqnarray}

Furthermore, it preserves $dX_h$, the harmonic map on $T^2$ given by (\ref{ec2notas}), and thus it is a symmetry of the Hamiltonian and of the full mass operator. This set of transformations is a generalization of the monodromy-generated invariance on an orbit of charges ($gQ\subset C_F$).

 It can be seen that the matrix $\mathbb{W}'$ has
 $det(\mathbb{W}')=det(\mathbb{W})=n\in \mathbb{Z}$. Although $\mathbb{W}'$ has, in general, noninteger elements, as it happens in the case of wrapped open string or in the case of free acting orbifolds, it preserves the central charge related to the winding term on the mass operator.

Consequently, the mass operator (\ref{MassOp_monodromy}) is invariant on the equivalence class of charges given by the coinvariant for a given parabolic, elliptic or hyperbolic monodromy.

Furthermore, the charges on the same coinvariant define the same symplectic torus bundle and hence the same physical M2-brane with monodromy. Each M2-brane bundle with a given pair of charges has moduli $R$ and $\tau$, that transform accordingly under the action of the U-duality group.  Different charges of the same coinvariant are associated to different moduli related among them by (\ref{transtau}) and (\ref{transR}). The mass operator of the M2-brane with an $SL(2,Z)$ monodromy is then expressed in terms of the coinvariant classes of KK charges and winding numbers. We interpret this symmetry as the origin of the gauge symmetry of the type IIB gauge supergravity in 9D. In order to understand better this relation we study the algebraic structure of the coinvariants.

\subsection{Algebraic structure of the coinvariants}

The new $SL(2,\mathbb{R})$ symmetry of the M2-brane with fluxes and monodromy, preserves the quantization of charges since it maps coinvariants onto coinvariants. We show that a coinvariant can be generated as an orbit of charges when specific matrices $\Lambda$ act on them
\begin{eqnarray}
  C_F = \Lambda \mathcal{Q}
\end{eqnarray}
where $C_F= \left (\begin{array}{c}
     p'  \\
     q'
\end{array}\right)$ and $\mathcal{Q}= \left (\begin{array}{c}
     p  \\
     q
\end{array}\right)$.
However, in general, the $\Lambda$ matrices does not form a group. Nevertheless, they are matrices of $SL(2,\mathbb{R})$, so they are conjugate to one of the one-parameter subgroups of $SL(2,\mathbb{R})$. Let us see case by case.

\paragraph{Parabolic case}
In this case,  from (\ref{CFP}) we have that $p'=p+\mathbb{Z}$ and $q'=q$. Consequently, from (\ref{Aoflambda}) we get that there are two possible values of $a$ such that $\Lambda$ is parabolic
\begin{eqnarray}
    \left\lbrace \begin{array}{c}
         a=  1\, , \, \mbox{ for } \, \, (\mbox{Tr}(\Lambda))=2 \\
         a=  \frac{-4p-3\mathbb{Z}}{\mathbb{Z}} \, , \, \mbox{ for } \, \, (\mbox{Tr}(\Lambda))=-2
    \end{array}\right.
\end{eqnarray}
 and from (\ref{lambda_SL(2,Q)}) the corresponding $\Lambda$ matrices that map coinvariants onto coinvariants are given by
 \begin{eqnarray}
     \Lambda_p^{(+2)} &=& \left( \begin{array}{cc}
        1  & \frac{\mathbb{Z}}{q}  \\
         0 & 1
     \end{array}\right) , \\
     \Lambda_p^{(-2)} &=& \left( \begin{array}{cc}
        -\frac{4p}{\mathbb{Z}}-3 & \frac{4p}{q}\frac{p+\mathbb{Z}}{\mathbb{Z}}+\frac{\mathbb{Z}}{q}  \\
         -\frac{4q}{\mathbb{Z}} & 1+\frac{4p}{\mathbb{Z}}
     \end{array}\right)
 \end{eqnarray}
It is easy to see that the case where $Tr(\Lambda)=+2$ is the only case where the $\Lambda$ matrices form a group. This is a parabolic subgroup and this is the case that was previously studied in \cite{mpgm23}. It means that one is allowed to move within the $\mathbb{Z}$ subclasses of the same coinvariant by consecutive transformations of $\Lambda$ matrices, and those transformations form a parabolic subgroup.

\paragraph{Elliptic case}
In the elliptic case, the monodromies are classified in conjugacy classes of matrices with trace $-1$, $0$ and $1$ isomorphic to $Z_3$, $Z_4$ and $Z_6$, respectively. In terms of the $\Lambda$ matrices, there are infinite numbers of $a$ values such that $-2<Tr(\Lambda_e)<2$.

For monodromies $\mathcal{M}_{Z_k}$ given by (\ref{ellipticmonodromyUV}) with $k=3,4,6$ we have from (\ref{CFE}) that $p'=p+\mathbb{Z}$ and $q'=q+\mathbb{Z}-k\hat{q}$. As it was already mentioned, the number of coinvariants is finite. There are $k$ coinvariants with $k$ subclasses each.

The $\Lambda_e$ matrices from (\ref{lambda_SL(2,Q)}) have restricted values of $a$, which are given by
\begin{eqnarray}
    a = 1 + \frac{(Q_e-2)q(p+\mathbb{Z})-\mathbb{Z}(\mathbb{Z}-k\hat{q})}{q\mathbb{Z}-p(\mathbb{Z}-k\hat{q})}.
    \end{eqnarray}
    where $Q_e=Tr(\Lambda_e)$. Even though these matrices $\Lambda_e$ transform coinvariants onto coinvariants, they do not form a group but a grupoid, because the clausure condition is not respected. 
    
    Nevertheless, from the Appendix \ref{ApeA}  we can see that $\Lambda_e=U^{-1}g_{SO(2)}U$ where $g$ is given by (\ref{gSO(2)2}) and the elements of the matrix $U\in SL(2,\mathbb{R})$ are given by 
\begin{eqnarray}
    U_{11} &=& \pm \left\lbrace \frac{\pm \sqrt{4-Q_e^2}}{\frac{2}{p+\mathbb{Z}}\left[(q+\mathbb{Z}-k\hat{q})\left(1+a\right)-q \right]}\right\rbrace ^{1/2}\\
    U_{12} &=& \pm \frac{a - \frac{1}{q}\left[(q+\mathbb{Z}-k\hat{q})-\frac{p(q+\mathbb{Z}-k\hat{q})}{p+\mathbb{Z}}a+\frac{qp}{p+\mathbb{Z}}\right]}{\sqrt{\pm\sqrt{4-Q_e^2}}\frac{2}{p+\mathbb{Z}}\left[(q+\mathbb{Z}-k\hat{q})\left(1+a\right)-q \right]} \\
    U_{21} &=& 0 \\
    U_{22} &=& \pm \left\lbrace \frac{\pm \sqrt{4-Q_e^2}}{\frac{2}{p+\mathbb{Z}}\left[(q+\mathbb{Z}-k\hat{q})\left(1+a\right)-q \right]}\right\rbrace ^{-1/2}
\end{eqnarray}

That is, each $\Lambda_e$ matrix that allows us to establish a transformation from coinvariant onto coinvariant is conjugated to an $SO(2)$ matrix, which can be written in terms of the trace $Q_E$ as in (\ref{gSO(2)}). Moreover, if we identify each of the $k$ coinvariants with $\theta_n$ with $n=1,\dots,k$, then the $SO(2)$ matrix can be written as (\ref{gSO(2)2}) with $\varphi=\theta_n-\theta_{m}=(n-m)\left( \frac{2\pi}{k} \right)$ with $m=1,\dots,n$.

Each value of $\Delta\theta$ will be related with a particular value of $Q_e=\mbox{Tr}(\Lambda_e)$. For $k=3$, the transformation within subclasses is done with elliptic matrices of $Q_e=-1$, for $k=4$ with matrices of $Q_e=-2,0$ and for $k=6$ with matrices of $Q_e=-2,-1,1$. In all the cases, the transformation that leaves the subclass invariant is given by $\Delta\theta=0$ and the corresponding trace is $Q_e=2$. Notice that, although we are considering elliptic matrices ($-2<Q_e<2$) of $SL(2,\mathbb{R})$, they are conjugate to $SO(2)$ as it is shown in the appendix. Therefore, the allowed rotations proportional to the identity are related with the critical cases $Q_e=\pm 2$.

Consequently, for an elliptic monodromy $\mathcal{M}_{Z_k}$, with $k=3,4,6$, we have that the parameter of $SO(2)$ parametrizing the transformation of an elliptic coinvariant onto itself is restricted to multiples of $\frac{2\pi}{k}$. Each discrete value of the parameter is related to an integer trace of the $\Lambda$ matrix. Here, we observe the origin at quantum level of the elliptic gauging. 

\paragraph{Hyperbolic case} 
In the hyperbolic case, the monodromies are classified in conjugacy classes of matrices with trace $\abs{n}\geq 3$. In terms of the $\Lambda$ matrices, there are in infinity of values of $a$ such that $\abs{Tr(\Lambda_h)}>2$. 

For monodromies $\mathcal{M}_h$ given by (\ref{hiperbolicmonodromyUV}), we have from (\ref{CFH}) that $p'=p+(n-2)\hat{p}+\mathbb{Z})$ and $q'=q+\mathbb{Z}$. As it was already mentioned there are $(n-2)$ coinvariants with $(n-2)$ subclasses each for a given $n$.

The $\Lambda_h$ matrices can be seen from (\ref{lambda_SL(2,Q)}) with $a$ given by
\begin{eqnarray}
    a = 1 + \frac{(Q_h-2)q(p+(n-2)\hat{p}+\mathbb{Z})-\mathbb{Z}((n-2)\hat{p}+\mathbb{Z})}{q((n-2)\hat{p}+\mathbb{Z})-p\mathbb{Z})}.
    \end{eqnarray}
    with $Q_h=Tr(\Lambda_h)$. Although these matrices $\Lambda_H$ transform coinvariants onto coinvariants, as happens in the elliptic case they do not form a group because the clausure condition is not respected. 
    
    Nevertheless, from the Appendix \ref{ApeA} we observe  that $\Lambda_H=U^{-1}g_{SO(1,1)}U$ where $g$ is given by (\ref{gSO(1,1)2}) and $U\in SL(2,\mathbb{R})$ is given by 
\begin{eqnarray}
    U_{11} &=& \pm  \left[Q_h - \frac{2}{q}\left( q+\mathbb{Z} - \frac{p(q+\mathbb{Z})}{p+(n-2)\hat{p}+\mathbb{Z}}a+\frac{qp}{p+(n-2)\hat{p}+\mathbb{Z}}\right)\pm \sqrt{Q_h^2-4}\right], \nonumber \\ 
    &*&\left[\pm \frac{\hat{D}}{D\sqrt{Q_h^2-4}} + \frac{D}{4\hat{D}}\frac{p+(n-2)\hat{p}+\mathbb{Z}}{(q+\mathbb{Z})a-q} \right] \\
    U_{12} &=& \mp  \left[Q_h - \frac{2}{q}\left( q+\mathbb{Z} - \frac{p(q+\mathbb{Z})}{p+(n-2)\hat{p}+\mathbb{Z}}a+\frac{qp}{p+(n-2)\hat{p}+\mathbb{Z}}\right)\pm \sqrt{Q_h^2-4}\right], \nonumber \\
    &*& \left[ \frac{\hat{D}}{D\sqrt{Q_h^2-4}} \mp \frac{D}{4\hat{D}}\frac{p+(n-2)\hat{p}+\mathbb{Z}}{(q+\mathbb{Z})a-q} \right] \\
    U_{21} &=& \frac{2\hat{D}}{D\sqrt{Q_h^2-4}}\frac{1}{p+(n-2)\hat{p}+\mathbb{Z}}\left[(q+\mathbb{Z})a - q\right] \pm \frac{\hat{D}}{2D} \\
    U_{22} &=& \mp \frac{2\hat{D}}{D\sqrt{Q^2-4}}\frac{1}{p+(n-2)\hat{p}+\mathbb{Z}}\left[(q+\mathbb{Z})a - q\right] + \frac{\hat{D}}{2D}
\end{eqnarray}
with $D,\hat{D}\in\mathbb{R}$. 

That is, each $\Lambda_h$ matrix that allows to establish a transformation from the hiperbolic coinvariant onto itself is conjugated to an $SO(1,1)$ matrix and it can be written in terms of the trace $Q_h$ as in (\ref{lambda_SL(2,Q)}). Moreover, if we identify each of the $(n-2)$ coinvariants with $\theta_r$, and $r=1,\dots,(n-2)$, then the $SO(1,1)$ matrix can be written as (\ref{gSO(1,1)2}) with $a=\theta_r-\theta_{s}=(n-3)\ln{\left(\frac{n}{2}+\sqrt{\frac{n^2}{4}-1}\right)}$ with $s=1,\dots,(n-2)$. 

Consequently, for  hyperbolic monodromies $\mathcal{M}_{h}$, we have that the paramenter of $SO(1,1)$ at quantum level become restricted to $(n-3)$ multiples of $\ln{\left(\frac{n}{2}+\sqrt{\frac{n^2}{4}-1}\right)}$ with $\abs{n}\geq 3$. For each discrete value of the parameter, there the trace of the matrix $\Lambda$ has an integer value given by
\begin{eqnarray}
    Q_h &=& \frac{\left( \frac{n}{2}+\frac{\sqrt{n^2-4}}{2}\right)^{2(n-3)}+1}{\left( \frac{n}{2}+\frac{\sqrt{n^2-4}}{2}\right)^{(n-3)}}
\end{eqnarray}

We observe that, for parabolic, elliptic or hyperbolic monodromies, the quantization of the values is dictated not only because of the quantization of charges as previously conjectured by Hull \cite{Hull4,Hull8}, but it is a consequence of the twisted torus bundle structure the M2-brane described in terms of coinvariants.

\subsection{Transformations between different coinvariants}

We are now considering a formulation of the twisted M2-brane in terms of the module of $\mathcal{M}_g$-coinvariants with $\mathcal{M}_g\subset SL(2,\mathbb{Z})$. It follows from the explicit expression of the mass operator that, indeed, it is defined on the coinvariant classes. 

Let us identify the transformations that relate inequivalent classes of M2-brane twisted torus bundles with monodromy. This is equivalent to determining the transformation that relates the different coinvariant classes associated with $\mathcal{M}_g$. It turns out that this transformation is a symmetry of the formulation if we consider the corresponding transformation on the moduli and the charges (See \cite{mpgm23} for the parabolic case). If the monodromy is trivial, each pair of charges ($p,q$) represents a coinvariant and the symmetry of the formulation is $SL(2,\mathbb{Z})$ as determined by \cite{Schwarz6}. For a nontrivial monodromy, the space of ($p,q$) points is distributed in terms of disjoint coinvariants associated with $\mathcal{M}_g$, and the M2-brane is a theory on the module of $\mathcal{M}_g$-coinvariants.

Following \cite{mpgm23}, we introduce some formal definitions that will allow us to determine the precise bundle coinvariant transformation. 
Given a group $G$ and a subgroup $H\in G$ we define the following classes
\begin{eqnarray}
    aH=\left\lbrace ah:h\in H \right\rbrace, a\in G, \label{classesequiv}
\end{eqnarray}
There is an equivalence relationship between two elements $a,b\in G$ provided that $b=ah$ for some $h\in H$. This relation can be re-expressed as  $a=bh^{-1}$.

A relevant property is that each element $c\in G$ is contained in one and only one equivalence class. If $c=ah=b\widehat{h}\rightarrow a=b\widehat{h}h^{-1}\in bH$ and then the classes $aH=bH$. Hence $G$ is the disjoint union of the equivalence classes generated by the subgroup $H$.

Given any pair of charges 
$Q=
\left(\begin{array}{c}
     p  \\
     q 
\end{array}\right)
$
with $p,q\in \mathbb{Z}$
and $Q_0=\left(\begin{array}{c}
     1  \\
     0 
\end{array}\right)$, there exists a matrix $V\in GL(2,Z)$, such that
$
Q=VQ_0.
$ The most general expression preserving the determinant is
\begin{eqnarray}\label{genexpdet}
    \left(\begin{array}{cc}
    p & r + \lambda p'  \\
    q & s + \lambda q'
\end{array} \right) = \left(\begin{array}{cc}
    p & r  \\
    q & s
\end{array} \right)\left(\begin{array}{cc}
    1 & \frac{\lambda}{m}  \\
    0 & 1
\end{array} \right),
\end{eqnarray}
with $r$ and $s$ unique, $\lambda,m\in\mathbb{Z}$ such that $p=mp'$ and $q=mq'$ with $p',q'$ relatively primes. 

Let us define $\mathcal{V}$ as a linear representation of the discrete subgroup $\mathcal{M}_g$. The quotient $\frac{Q}{g\widehat{Q}-\widehat{Q}}$ is the module of $\mathcal{M}_g$-coinvariants \cite{Sharifi}. Two classes $C_{F_1}=\left\lbrace Q_1 + g\widehat{Q}-\widehat{Q} \right\rbrace$ and $C_{F_2}=\left\lbrace Q_2 + g\widehat{Q}-\widehat{Q} \right\rbrace$ are disjoint if and only if $Q_1$ and $Q_2$ are not in the same coinvariant. In these expressions $\widehat{Q}$ is an arbitrary element of the space $\mathcal{V}$ and $g$ any element of the subgroup $\mathcal{M}_g$.

In order to transform
\begin{eqnarray}
    C_{F_1}\rightarrow C_{F_2},
\end{eqnarray}
we perform the following transformation
\begin{eqnarray}
    C_{F_1}\xrightarrow[]{\Lambda_{1}^{-1}} Q_1= \left(\begin{array}{c}
     p_1  \\
     q_1 
\end{array}\right) \rightarrow Q_0=\left( \begin{array}{c}
         1  \\
         0 
    \end{array}\right)\rightarrow Q_2=\left(\begin{array}{c}
     p_2  \\
     q_2 
\end{array}\right) \xrightarrow[]{\Lambda_{2}} C_{F_2},
\end{eqnarray}
where $\Lambda_{1}$, $\Lambda_{2}$ are $SL(2,\mathbb{R})$ transformation within the same coinvariant. 

Following (\ref{classesequiv}), where $G=GL(2,Z)$ and $H=\mathcal{M}_g$, we have that (See \cite{mpgm23})
\begin{eqnarray}
   Q_1 \xrightarrow[]{ba^{-1}} Q_2.
\end{eqnarray}
where
\begin{eqnarray}
    a&=&\left(\begin{array}{cc}
     p_1 & r_1\\
     q_1 & s_1
\end{array}\right), \\
b&=&\left(\begin{array}{cc}
         p_2 & r_2  \\
         q_2 & s_2
    \end{array}\right)
\end{eqnarray}
Given $p_1,q_1$ ($p_2,q_2$), the most general expression for $a$ ($b$), preserving its determinant has the general expression (\ref{genexpdet}), with $a,b$ determined uniquely by $p_1,q_1$ and $p_2,q_2$ , respectively. 

It can be seen that the coinvariants for a given (parabolic, elliptic or hiperbolic) monodromy contain the element $ \left(\begin{array}{c}
         1  \\
          q
    \end{array}\right)$,
where the allowed values for $q$ and the element associated with the trivial class will vary from case to case.

Then, is easy to verify that 
\begin{eqnarray}
    \left(\begin{array}{c}
         1  \\
          q_2
    \end{array}\right) =\mathcal{M}_\beta  \left(\begin{array}{c}
         1  \\
          q_1
    \end{array}\right)
\end{eqnarray}
with 
\begin{eqnarray}
    \mathcal{M}_\beta \equiv ba^{-1} = \left(\begin{array}{cc}
        1 & 0  \\
        \beta & 1
    \end{array}\right)
\end{eqnarray}
where the allowed values of $\beta=q_2-q_1$ are restricted in each case, $  a= \left(\begin{array}{cc}
        1 & 1  \\
        q_1 & (q_1+1)
    \end{array}\right)$ and $b = \left(\begin{array}{cc}
        1 & 1  \\
        q_2 & (q_2+1)
    \end{array}\right),
$
with $q_2\neq q_1$. 

Therefore
\begin{eqnarray}
    \mathcal{M}_\beta= \left(\begin{array}{cc}
        1 & 0  \\ 
        1 & 1
    \end{array}\right)^\beta, \label{Mbsubgroup}
\end{eqnarray}
with the corresponding allowed values for $\beta$ in each case, will be isomorphic to a subgroup of $SL(2,\mathbb{Z})$.

As found in \cite{mpgm23} it can be seen that the transformation between coinvariants is given by
\begin{eqnarray}
    C_{F_1}\xrightarrow[]{\Lambda_{1}^{-1}} \left(\begin{array}{c}
     1  \\
     q_1 
\end{array}\right) \xrightarrow[]{\mathcal{M}_\beta}  \left(\begin{array}{c}
     1  \\
     q_2 
\end{array}\right) \xrightarrow[]{\Lambda_{2}} C_{F_2}.
\end{eqnarray}
Let us emphasize that this transformation maps integer charges into integer charges. In the following we will obtain the particular symmetry transformations allowed between different coinvariant class for the monodromies contained on $SL(2,\mathbb{Z})$.

\paragraph{Parabolic monodromy}
For parabolic monodromies the inequivalent classes of twisted torus bundles are given by the coinvariants (\ref{CFP}). We have $\mathbb{Z}$ coinvariants classified by the value of $q$ and  $\mathbb{Z}$ elements within each coinvariant. In this case $q\in \mathbb{Z}$ and consequently $\beta\in\mathbb{Z}$. The trivial class will correspond with $q=0$. It can be seen that the transformation between inequivalent coinvariants (\ref{Mbsubgroup}) with $\beta\in\mathbb{Z}$ corresponds to the generator of the parabolic subgroup of $SL(2,\mathbb{Z})$
\begin{eqnarray}
    \mathcal{M}_\beta= \left\lbrace 
    \left(\begin{array}{cc}
        1 & 0  \\ 
        0 & 1
    \end{array}\right), \left(\begin{array}{cc}
        1 & 0  \\ 
        1 & 1
    \end{array}\right), \dots, \left(\begin{array}{cc}
        1 & 0  \\ 
        \mathbb{Z} & 1
    \end{array}\right)  \right\rbrace
\end{eqnarray}

\paragraph{Elliptic monodromy}

From the elliptic coinvariants (\ref{CFE}), we know that there are $k$ inequivalent classes of twisted torus bundles for monodromies isomorphic to $\mathcal{Z}_k$ with $k=3,4,6$ with $k$ subclasses within each coinvariant. It can be checked that, the values of $q$ are restricted in this case to $q=\mathbb{Z} \mbox{mod}(k)$, and consequently $\beta=\mathbb{Z} \mbox{mod}(2k-3)$. The trivial class will correspond with $q=1\mbox{mod}(k)$ for each case. It can be seen that the transformation between inequivalent coinvariants (\ref{Mbsubgroup}) with $\gamma=\mathbb{Z} \mbox{mod}(2k-3)$, corresponds to the generator of cyclic group $\mathcal{Z}_{2k-3}$
\begin{eqnarray}
    \mathcal{M}_\beta &=& \left\lbrace 
    \left(\begin{array}{cc}
        1 & 0  \\ 
        0 & 1
    \end{array}\right), \left(\begin{array}{cc}
        1 & 0  \\ 
        1 & 1
    \end{array}\right) ,  \dots, \left(\begin{array}{cc}
        1 & 0  \\ 
        2k-3 & 1
    \end{array}\right) \right\rbrace, \nonumber \\
    &=& \left\lbrace 
   \left(\begin{array}{cc}
        1 & 0  \\ 
        -(k-2) & 1
    \end{array}\right),\dots, \left(\begin{array}{cc}
        1 & 0  \\ 
        0 & 1
    \end{array}\right), \dots, \left(\begin{array}{cc}
        1 & 0  \\ 
        (k-2) & 1
    \end{array}\right) \right\rbrace
\end{eqnarray}
Although the transformation between inequivalent classes is isomorphic to the cyclic group of $2k-3$ elements, there is a unique homeomorphism with the cyclic group $\mathcal{Z}_{(k-1)}$
\begin{eqnarray}
    \mathcal{M}_\beta 
    &=& \left\lbrace \left(\begin{array}{cc}
        1 & 0  \\ 
        0 & 1
    \end{array}\right), \dots, \left(\begin{array}{cc}
        1 & 0  \\ 
        (k-2) & 1
    \end{array}\right) \right\rbrace
\end{eqnarray}

\paragraph{Hyperbolic monodromies}
From the hyperbolic coinvariant (\ref{CFH}), we know that there are $(n-2)$ inequivalent classes of twisted torus bundles for monodromies with $(n-2)$ subclasses within each coinvariant and $n\geq3$. It can be checked that, the values of $q$ are restricted in this case to $q=\mathbb{Z} \mbox{mod}(n-2)$, and consequently $\beta=\mathbb{Z} \mbox{mod}(2n-7)$. The trivial class will correspond with $q=1\mbox{mod}(n-2)$ for each case. It can be seen that the transformation between inequivalent coinvariants (\ref{Mbsubgroup}) with $\gamma=\mathbb{Z} \mbox{mod}(2n-7)$, it corresponds to the generator of cyclic group $\mathcal{Z}_{(2n-7)}$ for a given $n$ (greater than $2$)
\begin{eqnarray}
    \mathcal{M}_\beta &=& \left\lbrace 
    \left(\begin{array}{cc}
        1 & 0  \\ 
        0 & 1
    \end{array}\right), \left(\begin{array}{cc}
        1 & 0  \\ 
        1 & 1
    \end{array}\right) ,  \dots, \left(\begin{array}{cc}
        1 & 0  \\ 
        2n-7 & 1
    \end{array}\right) \right\rbrace, \nonumber \\
    &=& \left\lbrace 
   \left(\begin{array}{cc}
        1 & 0  \\ 
        -(n-4) & 1
    \end{array}\right),\dots, \left(\begin{array}{cc}
        1 & 0  \\ 
        0 & 1
    \end{array}\right), \dots, \left(\begin{array}{cc}
        1 & 0  \\ 
        (n-4) & 1
    \end{array}\right) \right\rbrace
\end{eqnarray}
Although hyperbolic monodromies requires $n\geq3$, in order for the transformation between inequivalent coinvariant make sense  it is neccessary to have $n\geq 5$. This is because the transformation only make sense between the nontrivial classes. For a given $n$, there are $(n-2)$ inequivalent coinvariants and $(n-3)$ nontrivial classes. 

As happens in the elliptic case, there is a unique homeomorphism to the cyclic group $\mathcal{Z}_{(n-3)}$ 
\begin{eqnarray}
    \mathcal{M}_\beta 
    &=& \left\lbrace \left(\begin{array}{cc}
        1 & 0  \\ 
        0 & 1
    \end{array}\right), \dots, \left(\begin{array}{cc}
        1 & 0  \\ 
        (n-4) & 1
    \end{array}\right) \right\rbrace
\end{eqnarray}

The transformation between inequivalent coinvariants for a given monodromy is realized as a discrete symmetry of the quantum M2-brane theory. It corresponds to the quantization of the $SL(2,\mathbb{R})$ to subgroups of the $SL(2,\mathbb{Z})$.

 Consequently, by taking into account the corresponding transformation on the moduli and the charges, we can define a transformation 
 related to $\widetilde{\Lambda}=\Lambda_{2}\mathcal{M}_\beta {\Lambda_1}^{-1}$, as described in \cite{mpgm23}, that leaves invariant the M2-brane mass operator. It can be interpreted as a duality between inequivalent classes of M2-brane twisted torus bundles for a given monodromy.

\section{Conclusions}

We have established a relation between M2-branes with nonvanishing winding on a torus and nontrivial monodromy and the type IIB $\mathbb{R}$, $SO(2)$ and $SO(1,1)$ gauged supergravities in nine dimensions. The global description of those M2-branes is given by twisted torus bundles with monodromy in $SL(2,\mathbb{Z})$. Inequivalent classes of torus bundles, for a given monodromy, are given by the coinvariants. We find that the group transformation between equivalent bundles coincides with the symmetry group of the corresponding gauge supergravity. The discretization of the mass parameters which are consistent with the quantum theory are those related with the subclasses of charges within each coinvariant. On the other hand, the group transformation between inequivalent twisted torus bundles, is related with the U-duality subgroup that acts as a global symmetry on the theory. While in the quantum realization of the maximal supergravity case, the U-duality corresponds to the discrete subgroup $SL(2,\mathbb{Z})$ of $SL(2,\mathbb{R})$, in the case of the type II gauged supergravity cases we find that it corresponds to discrete subgroups of $SL(2,\mathbb{Z})$ as it can be seen from table (\ref{table}).
We first noticed that the symplectic connection of the M2-brane worldvolume formulation, transforms as the gauge vector for all type II gauge supergravities in nine dimensions. This suggest a deeper relation to be explored in the future.\newline
Originally, the worldvolume description of the M2-brane with nonvanishing winding on a torus and nontrivial monodromy is formulated in an orbit of charges contained in the coinvariant $gQ\subset C_F$. We showed in \cite{mpgm23} that, for parabolic monodromies, this symmetry can be extended such that the local description can be formulated on the coinvariants that classify  the global theory. In this work, we generalize this result for the elliptic and hyperbolic monodromies. In this last two cases we show that the Hamiltonian of the M2-brane with nonvanishing winding on a torus and nontrivial monodromy, is formulated on the coinvariants. For the parabolic and the hyperbolic case, there exist an infinite number of coinvariants, while for the elliptic scenario, as expected, the inequivalent coinvariants are finite. 

In order to formulate the M2-brane on the coinvariants for a given monodromy, the symmetry of the mass operator is extended to $SL(2,\mathbb{R})$, where the generator of the transformation are parabolic, elliptic o hyperbolic matrices, according to the monodromy. This symmetry is not present in the case with trivial monodromy where eaych coinvariant contains only one element. When the monodromy is nontrivial, we identify the symmetry relating the elements of a coinvariant as a gauge symmetry of the formulation. In fact, not only the physical content remains invariant but also the geometric formulation is defined on the same twisted torus bundles. The generator of the gauge symmetry for a given monodromy is, in general, conjugated to the corresponding one-parameter subgroup of $SL(2,\mathbb{R})$, i.e. $\mathbb{R}$, $SO(2)$ or $SO(1,1)$ for parabolic, elliptic or hyperbolic monodromies, respectively. These subgroups correspond with the symmetry group of the type IIB gauge supergravity in nine dimensions known as the triplet. The mass parameter is compact in the elliptic case, and noncompact in the parabolic or hyperbolic subgroups. At low energies, all possible values are allowed in each case. Nevertheless, only \textit{particular} discrete values will be consistent with the quantum theory. We show that these values are related with the identification of coinvariants for a given monodromy. Consequently, in this paper we have found new evidences that show that the low energy limit of the M2-brane with nonvanishing winding and parabolic, elliptic or hyperbolic monodromy is given by the type IIB $\mathbb{R}$, $SO(2)$ or $SO(1,1)$ gauge supergravity in nine dimensions, respectively.

We also demonstrate that the transformation between M2-brane twisted torus bundles with parabolic monodromy but different second cohomology class, i.e. different coinvariants, can be expressed in terms of a subgroup $\mathcal{M}_\beta$ where the allowed values of $\beta$ will be restricted for each monodromy. For parabolic monodromies, $\beta\in \mathbb{Z}$ and $\mathcal{M}_\beta$ is conjugated to $\mathcal{M}_p$, the parabolic subgroup of $SL(2,\mathbb{Z})$. For elliptic monodromies $\mathcal{Z}_k$ with $k=3,4,6$ we have that $\beta=\mathbb{Z}\mbox{mod}(2k-3)$ and $\mathcal{M}_\beta$ corresponds to the generator of $\mathcal{Z}_{2k-3}$, the cyclic group of $(2k-3)$ elements. Nevertheless, although the group of transformations between different coinvariants is isomorphic to $\mathcal{Z}_{2k-3}$, is also homemorphic to $\mathcal{Z}_{k-1}$, the cyclic group of $(k-1)$ elements. This can be understood as the identification of a matrix and it inverse to the same element in the homemorphism, and it can be also interpreted from the M2-brane coinvariants by allowing transformations in the same direction. For hiperbolic monodromies, it can be seen that $\beta=\mathbb{Z}\mbox{mod}(2n-7)$, with $n\geq 5$ and $\mathcal{M}_\beta$ corresponds to the generator of $\mathcal{Z}_{2n-7}$, the cyclic group of $(2n-7)$ elements. Nevertheless, although the group of transformations between different coinvariants is isomorphic to $\mathcal{Z}_{2n-7}$, is also homemorphic to $\mathcal{Z}_{n-3}$, the cyclic group of $(n-3)$ elements. For hiperbolic monodromies $n>2$ in general, but it can checked that the transformation between inequivalent coinvariants only make sense for $n\geq 5$.
\begin{table}
\label{table}
\begin{tabular}{|c|c|c|c|c|}
     \hline
    Monodromy & $\mathcal{M}_g\subset SL(2,\mathbb{Z})$ & Gauge Group  & Quantum restriction & U-duality group  \\
 \hline
    Trivial & $\mathbb{I}$ & - & $SL(2,\mathbb{Z})$ & $SL(2,\mathbb{Z})$ \\
    \hline
    Parabolic & $\mathcal{M}_p$ & $\mathbb{R}$ & $\mathbb{Z}$ & $\mathbb{Z}$ \\
    \hline
       Elliptic & $\mathcal{M}_{\mathcal{Z}_3}$ & $SO(2)$ & $\mathcal{Z}_3$ & $\mathcal{Z}_3$ \\
        \hline
      Elliptic & $\mathcal{M}_{\mathcal{Z}_4}$ &  $SO(2)$ & $\mathcal{Z}_4$ & $\mathcal{Z}_5$\\ 
        \hline
      Elliptic & $\mathcal{M}_{\mathcal{Z}_6}$ &  $SO(2)$ & $\mathcal{Z}_6$ & $\mathcal{Z}_9$\\
        \hline
      Hiperbolic & $\mathcal{M}_{n}$ &  $SO(1,1)$ & -  & $\mathcal{Z}_{(2n-7)}$\\ 
      \hline
\end{tabular}    
\caption{Gauge group, discrete values of the mass parameters and U-duality groups of type IIB gauge supergravities from M2-brane twisted torus bundles}
\end{table}
For the case with trivial monodromy, the coinvariants contains only one element and the transformation between different coinvariants is given by $SL(2,\mathbb{Z})$, the U-duality group. Consequently, we claim that for the cases with nontrivial monodromy, the transformation between different coinvariants will corresponds to the U-duality group for a given monodromy. It is always related with discrete subgroups of $SL(2,\mathbb{Z})$. The U-duality group is $\mathbb{Z}$ for parabolic monodromies, $\mathcal{Z}_3$, $\mathcal{Z}_5$ y $\mathcal{Z}_9$ for the elliptic case and  $\mathcal{Z}_{(2n-7)}$ for hiperbolic monodromies.

The double dimensional reduction of these M2-branes with fluxes and monodromy will be considered in \cite{mpgm25}. This will be done following \cite{mpgm23} where parabolic ($p,q$)-strings on a circle were obtained from the M2-brane with nonvanishing winding on a torus and parabolic monodromy. We will also discuss some features of the swampland program for type II maximal supergravity in nine dimensions from the M2-brane worldvolume perspective. 

\appendix
\section{One parameter groups of $SL(2,\mathbb{R})$}.
\label{ApeA}

In this appendix we will review the one-parameter subgroups of $SL(2,\mathbb{R})$ and their application to the particular set of matrices considered in this work.

Let us consider a matrix 
\begin{eqnarray}
    A = \left(\begin{array}{cc}
        a & b \\
    c & d
    \end{array} \right)\in SL(2,\mathbb{R})
\end{eqnarray}

\begin{enumerate}
    \item If $Tr(A)=\pm 2$, we have that $A=UgU^{-1}$ where
    \begin{eqnarray}
        g &=& \left( \begin{array}{cc}
            \pm 1 & K \\
             0 &  \pm 1
        \end{array} \right) , \\
    U &=& \left( \begin{array}{cc}
        \frac{\pm 1 -d}{(\pm 1 -d)D -cB} & B   \\
         \frac{c}{(\pm 1 -d)D -cB} & D
    \end{array} \right) \in SL(2,\mathbb{R})
    \end{eqnarray}
with $K=2(\pm 1-d)DB - cB^2 + bD^2$. 

It can be seen that the matrices $g$ forms a group with the particularity the all matrices have $\abs{\mbox{Tr}(g)}=2$. It means that all parabolic matrices of $SL(2,\mathbb{R})$ are conjugate to triangular matrices.

\item If $Tr(A)< 2$, we have that $A=UgU^{-1}$ where
\begin{eqnarray}
        g &=& \frac{1}{2}\left( \begin{array}{cc}
            Tr(A) & \mp \sqrt{4-Tr(A)^2} \\
             \pm \sqrt{4-Tr(A)^2} &  Tr(A)
        \end{array} \right) , \label{gSO(2)}\\
    U &=& \left( \begin{array}{cc}
        \pm \left( \pm \frac{\sqrt{4-Tr(A)^2}}{2c} \right)^{1/2} & \pm \frac{a-d}{\left( \pm 2c \sqrt{4-Tr(A)^2} \right)^{1/2}}   \\
         0 & \pm \left( \pm \frac{\sqrt{4-Tr(A)^2}}{2c} \right)^{-1/2}
    \end{array} \right) \in SL(2,\mathbb{R}) \label{USO(2)}
    \end{eqnarray}
It can be seen that the matrices $g$ forms a group $SO(2)$. Indeed, if $Tr(A)=2\cos(\varphi)$ we have that
\begin{eqnarray}
    g = \left( \begin{array}{cc}
       \cos(\varphi)  & \mp \sin(\varphi)  \\
          \pm \sin(\varphi) & \cos(\varphi) 
    \end{array}\right) \in SO(2) \label{gSO(2)2}
\end{eqnarray}
We have then that all elliptic matrices of $SL(2,\mathbb{R})$ are conjugate to elements of $SO(2)$.

\item  If $Tr(A) > 2$, we have that $A=Wg_1W^{-1}$ where
\begin{eqnarray}
        g_1 &=& \left( \begin{array}{cc}
            \lambda & 0 \\
             0 &  \lambda^{-1}
        \end{array} \right) , \\
    U &=& \left( \begin{array}{cc}
        \frac{1}{D}\left( \pm 1+ \frac{Tr(A)-2d}{\sqrt{Tr(A)^2-4}}\right) & \frac{D}{2c}\left( Tr(A)-2d\pm\sqrt{Tr(A)^2-4}\right)  \\
         \frac{2c}{D\sqrt{Tr(A)^2-4}} & D
    \end{array} \right) \in SL(2,\mathbb{R}) \label{USO(1,1)}
    \end{eqnarray}
with $D\in\mathbb{R}$ and
\begin{eqnarray}
    \lambda &=& \pm \frac{1}{2}\left( Tr(A) + \sqrt{Tr(A)^2-4} \right), \\
    \lambda^{-1} &=& \pm \frac{1}{2}\left(Tr(A) -\sqrt{Tr(A)^2-4} \right). 
\end{eqnarray}
However, for the KAN decomposition of $SL(2,\mathbb{R})$ is enough to consider $\lambda>0$. 

It can be seen that these matrices $g_1$ form a group. Moreover, they are conjugate to the hiperbolic rotations. Indeed, it can be checked that $g_1=V^{-1}g_2V$ where
\begin{eqnarray}
        g_2 &=& \frac{1}{2}\left( \begin{array}{cc}
            Tr(A) & \mp \sqrt{Tr(A)^2-4} \\
             \mp \sqrt{Tr(A)^2-4} &  Tr(A)
        \end{array} \right) , \label{gSO(1,1)} \\
    U &=& \left( \begin{array}{cc}
       \frac{1}{2D} & \pm D   \\
         \mp \frac{1}{2D} & D
    \end{array} \right) \in SL(2,\mathbb{R})
    \end{eqnarray}
It can be seen that the matrices $g_2$ forms a group $SO(1,1)$. This group has two connected components given by $SO(1,1)^+$ and $SO(1,1)^-$, determined by the corresponding sign in $g_2$.

As before, KAN decomposition of $SL(2,\mathbb{R})$ invited us to consider the $SO(1,1)^+$ component. Consequently, if $Tr(A)=2\cosh(\theta)$ we have that
\begin{eqnarray}
    g_2 = \left( \begin{array}{cc}
       \cosh(a)  &  \sinh(a)  \\
         \sinh(a) & \cosh(a) 
    \end{array}\right) \in SO(1,1)^+ \label{gSO(1,1)2}
\end{eqnarray}

As $UV\in SL(2,\mathbb{R})$, we have that all hiperbolic matrices of $SL(2,\mathbb{R})$ are conjugate to elements of $SO(1,1)^+$.

\end{enumerate}

\acknowledgments
MPGM is partially supported by the MCI  Spanish Grant, 
PID2021-125700NB-C21 MCI Spanish Grant (“Gravity, Supergravity and Superstrings” (GRASS)). CLH is partially supported by ANID POSTDOCTORADO BECAS CHILE/2022-74220044. CLH is also grateful to Project PID2021-123017NB-I00, funded by MCIN/AEI/10.13039/501100011033. A.R thanks to SEM 18-02 project from U. Antofagasta.











\end{document}